\documentclass[apj]{emulateapj}

\usepackage{times}

\renewcommand{\ion}[2]{#1\,{\textsc{\romannumeral#2}}}

\newcommand{\hicLong}{High Resolution Coronal Imager}
\newcommand{\hic}{Hi-C}
\newcommand{\irisLong}{Interface Region Imaging Spectrograph}
\newcommand{\iris}{IRIS}
\newcommand{\rhessiLong}{The Reuven Ramaty High Energy Solar Spectroscopic Imager}
\newcommand{\rhessi}{RHESSI}
\newcommand{\xrt}{XRT}
\newcommand{\xrtLong}{X-Ray Telescope}
\newcommand{\eis}{EIS}
\newcommand{\eisLong}{Extreme Ultraviolet Imaging Spectrometer}
\newcommand{\aia}{AIA}
\newcommand{\aiaLong}{Atmospheric Imaging Assembly}
\newcommand{\hmi}{HMI}

\shorttitle{Signatures of Impulsive Heating Events. I. Observations}
\shortauthors{Warren et al.}

\slugcomment{To be submitted to the Astrophysical Journal}

\begin{document}


\title{Transition Region and Chromospheric Signatures of Impulsive Heating Events. I. Observations}

\author{Harry P. Warren\altaffilmark{1}, Jeffrey W. Reep\altaffilmark{2}, Nicholas
  A. Crump\altaffilmark{3}, Paulo J. A. Sim\~oes\altaffilmark{4}}

\altaffiltext{1}{Space Science Division, Naval Research Laboratory, Washington, DC 20375}
\altaffiltext{2}{National Research Council Postdoctoral Fellow, Space Science Division, Naval
  Research Laboratory, Washington, DC 20375}
\altaffiltext{3}{Naval Center for Space Technology, Naval Research Laboratory, Washington, DC 20375}
\altaffiltext{4}{SUPA School of Physics and Astronomy, University of Glasgow, Glasgow G12 8QQ, UK}


\begin{abstract}
  We exploit the high spatial resolution and high cadence of the \irisLong\ (\iris) to investigate
  the response of the transition region and chromosphere to energy deposition during a small
  flare. Simultaneous observations from \rhessi\ provide constraints on the energetic electrons
  precipitating into the flare footpoints while observations of \xrt, \aia, and \eis\ allow us to
  measure the temperatures and emission measures from the resulting flare loops. We find clear
  evidence for heating over an extended period on the spatial scale of a single
  \iris\ pixel. During the impulsive phase of this event the intensities in each pixel for the
  \ion{Si}{4}\,1402.770\,\AA, \ion{C}{2}\,1334.535\,\AA, \ion{Mg}{2} 2796.354\,\AA, and \ion{O}{1}
  1355.598\,\AA\ emission lines are characterized by numerous, small-scale bursts typically lasting
  60\,s or less. Red shifts are observed in \ion{Si}{4}, \ion{C}{2}, and \ion{Mg}{2} during the
  impulsive phase. \ion{Mg}{2} shows red-shifts during the bursts and stationary emission at other
  times. The \ion{Si}{4} and \ion{C}{2} profiles, in contrast, are observed to be red-shifted at
  all times during the impulsive phase.  These persistent red-shifts are a challenge for
  one-dimensional hydrodynamic models, which predict only short-duration downflows in response to
  impulsive heating. We conjecture that energy is being released on many small-scale filaments with
  a power-law distribution of heating rates.
\end{abstract}



\section{introduction}

Understanding how the solar upper atmosphere is heated to high temperatures is a fundamental
problem in solar physics. It has long been recognized that the solar chromosphere and transition
region, which supply the corona with mass, should provide important diagnostic information on the
energy release mechanism.  The complex topology and rapid evolution of these layers of the solar
atmosphere, however, has made this difficult to achieve in practice.

Recent observations from the \hicLong\ (\hic, \citealt{cirtain2013}) and the \irisLong\ (\iris,
\citealt{depontieu2014}) suggest that at sufficiently high spatial and temporal resolution it is
possible to track the response of the transition region and chromosphere to some individual heating
events. For example, \citet{testa2013} showed that at the approximately 150\,km spatial resolution
and 5.5\,s cadence of \hic, temporal variability in active region loop footpoints (the ``moss,''
e.g., \citealt{berger1999}) associated with the heating of some high temperature loops becomes
apparent. Lower resolution observations of the moss, in contrast, had suggested that the heating
was relatively steady \citep[e.g.,][]{antiochos2003, brooks2009}. Similarly, \citet{testa2014}
identified several events in 9.5\,s cadence \iris\ sit-and-stare observations that showed strong
blueshifts in \ion{Si}{4}. Numerical simulations indicate that these blueshifts are a signature of
energy deposition at heights below the region where \ion{Si}{4} is typically formed, perhaps
because the energy transport in these events is driven by electron beams rather than thermal
conduction. This recent work suggests that a detailed examination of the transition region and
chromosphere is likely to yield new insights into the physics of energy release during flares.

There is, of course, a long history of both observational and theoretical studies of impulsive
flare dynamics (for a review see \citealt{fletcher2011}). Previous observations have established
the close correspondence between the evolution of hard X-ray emission and emission from the
transition region and chromosphere \cite[e.g.,][]{kane1971, cheng1981, poland1982, woodgate1983,
  tandberg1983, hudson1994, simoes2015} as well as the presence of both evaporative upflows at high
temperatures \citep[e.g.,][]{doschek1980, antonucci1982} and downflows at transition region and
chromospheric temperatures \citep[e.g.,][]{ichimoto1984, zarro1988, canfield1990}.

Early numerical simulations were able to reproduce the very high coronal temperatures and densities
associated with impulsive energy deposition and chromospheric evaporation
\citep[e.g.,][]{nagai1980, fisher1985b, mariska1989}. These models were also able to account for
the red-shifts observed in the chromosphere and transition region, which are a consequence of the
sudden evaporative upflows at higher temperatures and momentum conservation \citep{fisher1989}. As
discussed by \citet{emslie1987}, however, these numerical simulations also predicted that the
evaporative upflows should dominate the velocity signature and the observed line profile should be
completely blueshifted. This is rarely observed in spatially unresolved observations
\citep{mariska1993}. At the spatial resolution of several thousand kilometers some completely
blueshifted high temperature profiles are observed \citep[e.g.,][]{czaykowska2001, doschek2013}. At
the approximately 200\,km spatial resolution of \iris, completely blueshifted \ion{Fe}{21} emission
is observed routinely \citep{tian2014, young2015, tian2015, graham2015, polito2016}. This
discrepancy is a consequence of the filamentary nature of energy release during a flare
\citep{hori1997,reeves2002,warren2006}.

In this paper we investigate the evolution of transition region and chromospheric line intensities,
velocities, and widths observed with a high cadence \iris\ sit-and-stare observation of a small
event (GOES class B4 after background subtraction) that occurred on 19 November 2014.  The
combination of data from many different satellites allows us to measure comprehensively the
properties and dynamics of the event in ways not possible with any individual instrument.  We
enumerate the constraints and observables that a model must be able to reproduce in order to
sufficiently understand the energy release.  This paper is part of a larger program to understand
the relationship between transition region emission and energy deposition in small events such as
microflares in the hope that these properties can be extrapolated to events that heat the solar
corona.
\section{observations}

\iris\ is a compact spectrograph based on a Cassegrain design. Special coatings allow for
simultaneous imaging of the 1332--1407 and 2783--2835\,\AA\ wavelength ranges. The far UV (FUV)
wavelength range includes strong emission lines from \ion{O}{1} 1355.598, \ion{C}{2} 1334.535 and
1335.708, \ion{Si}{4} 1393.755 and 1402.770, \ion{O}{4} 1399.775 and 1401.163, \ion{Fe}{12}
1349.382, and \ion{Fe}{21} 1354.080\,\AA. The near UV (NUV) wavelength range includes \ion{Mg}{2} k
2796 and \ion{Mg}{2} 2803\,\AA\ lines. The nearly 7\,m effective focal length provides a spatial
resolution scale of $0\farcs33$ or about 230\,km. Spectroscopy is provided by passing solar
radiation through a $0\farcs33\times175\arcsec$ slit and reflecting it off of a grating. The
resulting spectral resolution is about 26\,m\AA\ in the FUV and 53\,m\AA\ in the NUV. Light
reflected off of the slit assembly is passed through one of four science filters to allow for
context imaging of an area $175\arcsec\times175\arcsec$ around the slit. The high effective area of
\iris\ relative to previous spectrographs allows for much higher observing cadences, which are
typically below 10\,s.  \iris\ was launched into a sun-synchronous orbit and nearly continuous
observing is possible for about 9 months of the year. For additional details on the instrument see
\citet{depontieu2014}.

\begin{figure}[t!]
  \centerline{\includegraphics[bb=64 160 558 628,width=0.84\linewidth]{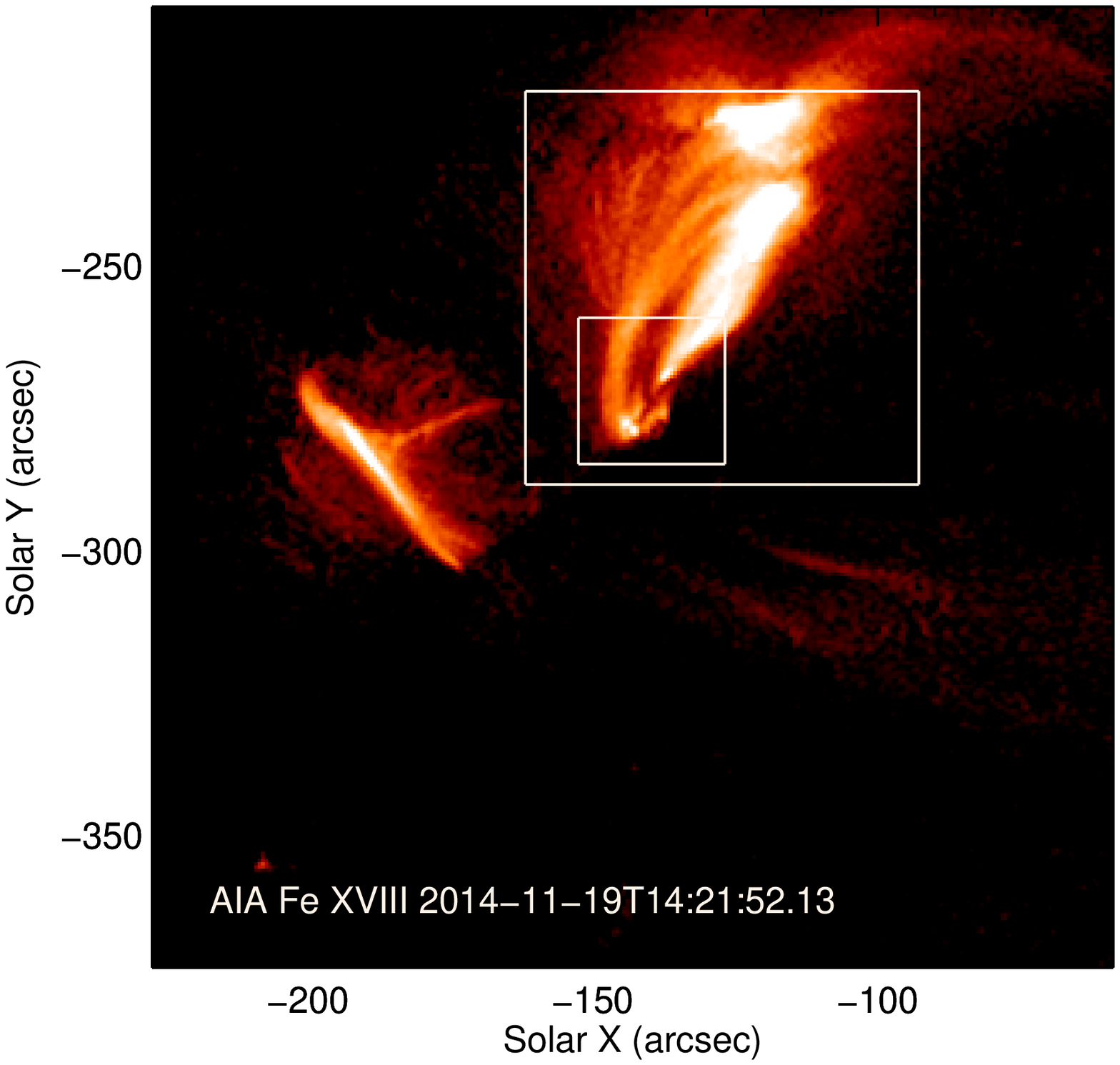}}
  \centerline{\includegraphics[bb=64 160 558 628,width=0.84\linewidth]{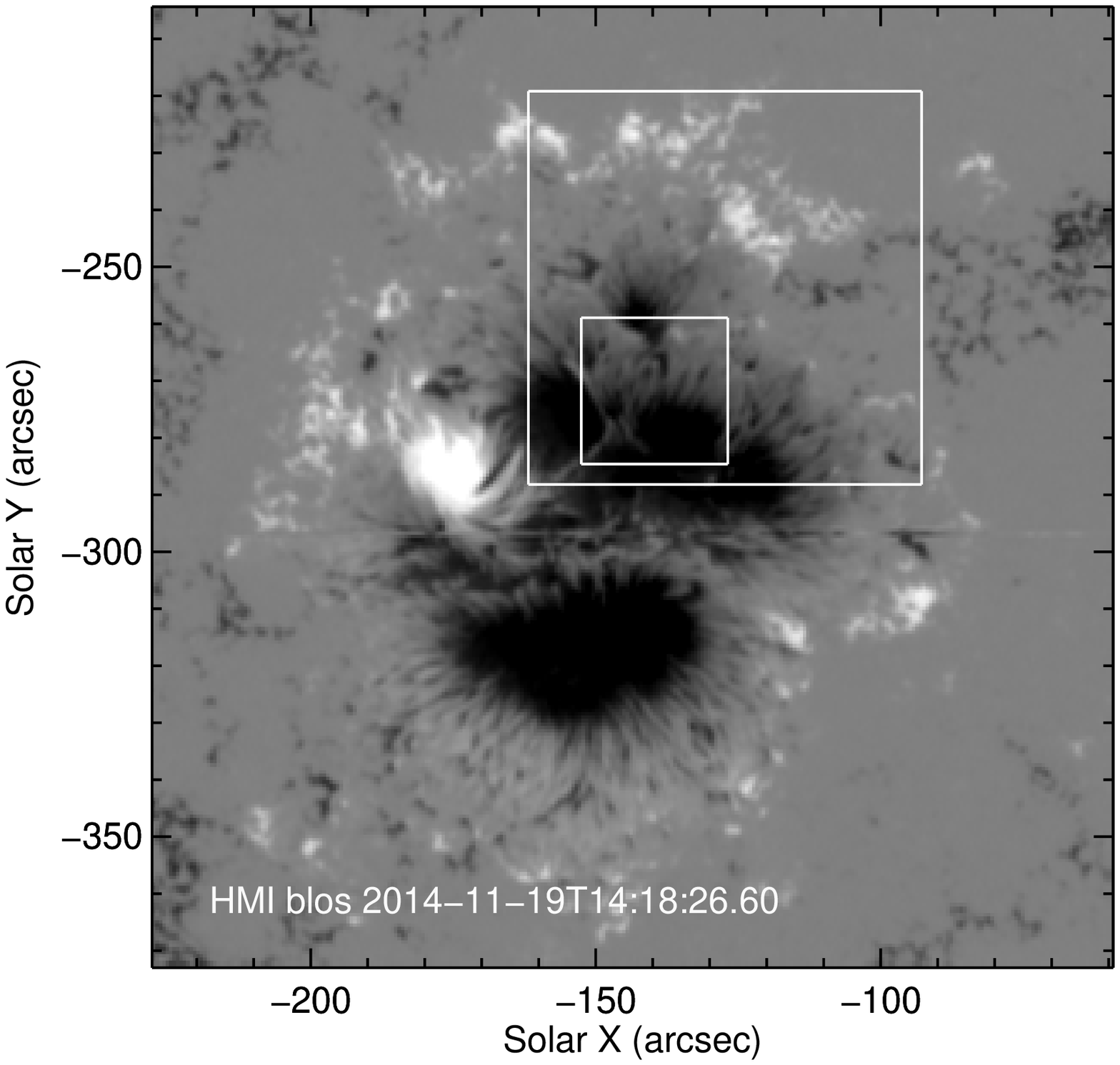}}
  \centerline{\includegraphics[bb=64 160 558 638,width=0.84\linewidth]{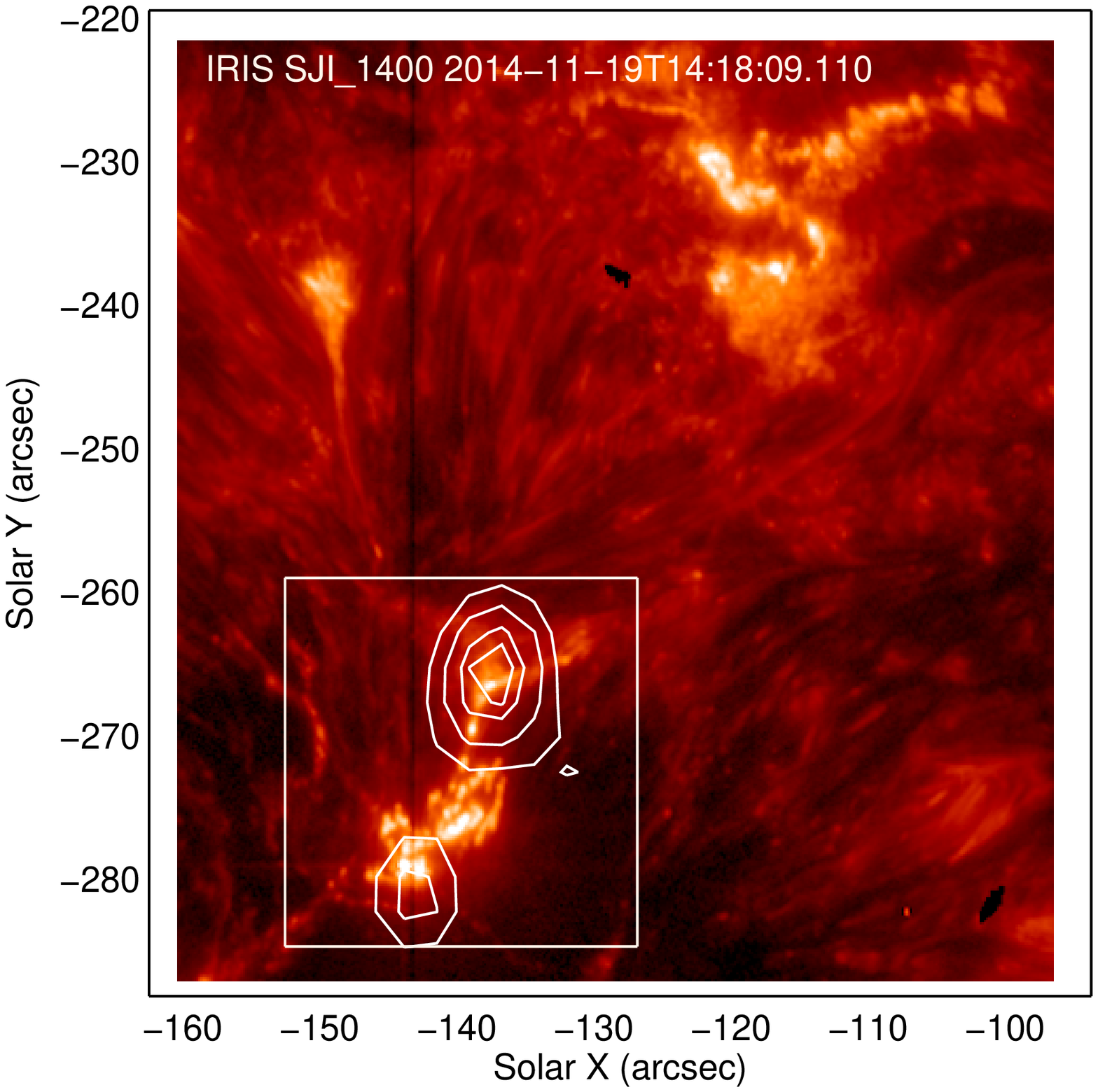}}
  \caption{\aia\ \ion{Fe}{18}, \hmi\ line-of-sight magnetogram, and \iris\ 1400\,\AA\ slit-jaw
    images taken near the peak of a small event, SOL2014-11-19T14:25\,UT. The peak GOES flux was
    about C1.1 and the event was not recorded as a flare on the GOES event
    list. \rhessi\ 15--25\,keV contours are shown on the \iris\ slit-jaw image. The dark vertical
    feature in the \iris\ slit-jaw image is the shadow of the slit. The small white box indicates
    the field of view shown in Figures~\ref{fig:rhessi}, \ref{fig:hist}, and \ref{fig:smallfov} and
    highlights the footpoint region near the sunspot penumbra.}
  \label{fig:context}
\end{figure}

\begin{figure*}[t!]
  \centerline{\includegraphics[width=\linewidth]{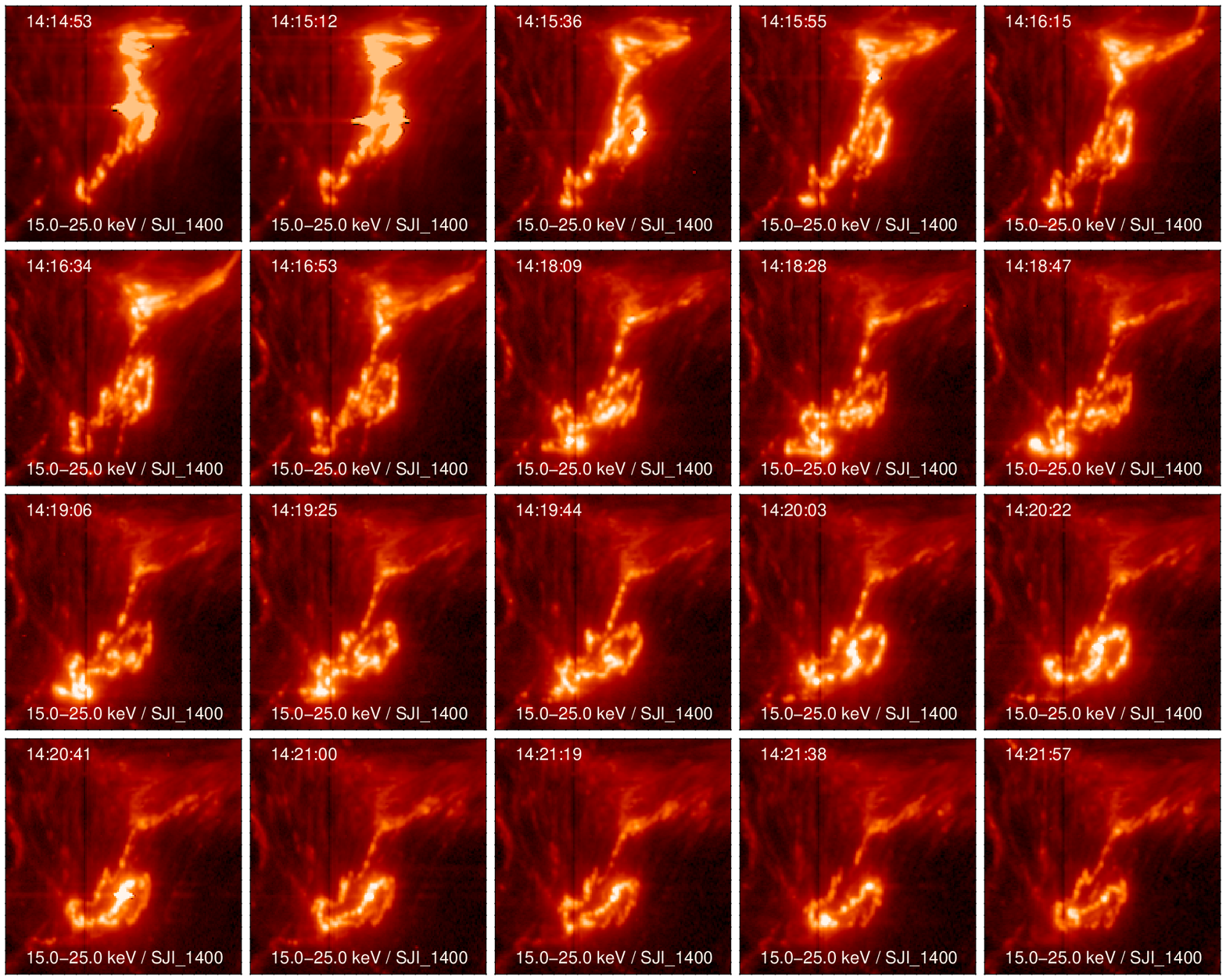}}
  \vspace{-0.1in}
  \centerline{\includegraphics[width=0.5\linewidth]{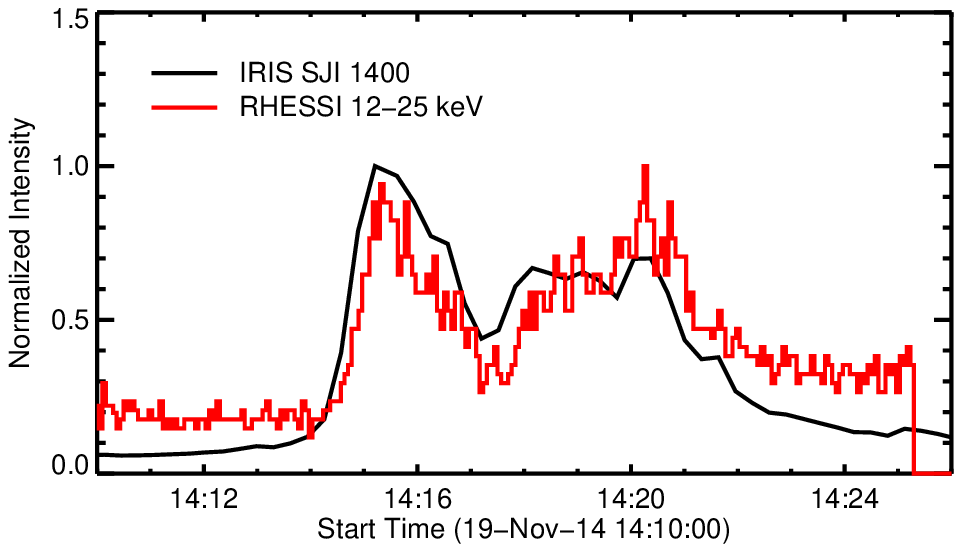}%
    \includegraphics[width=0.5\linewidth]{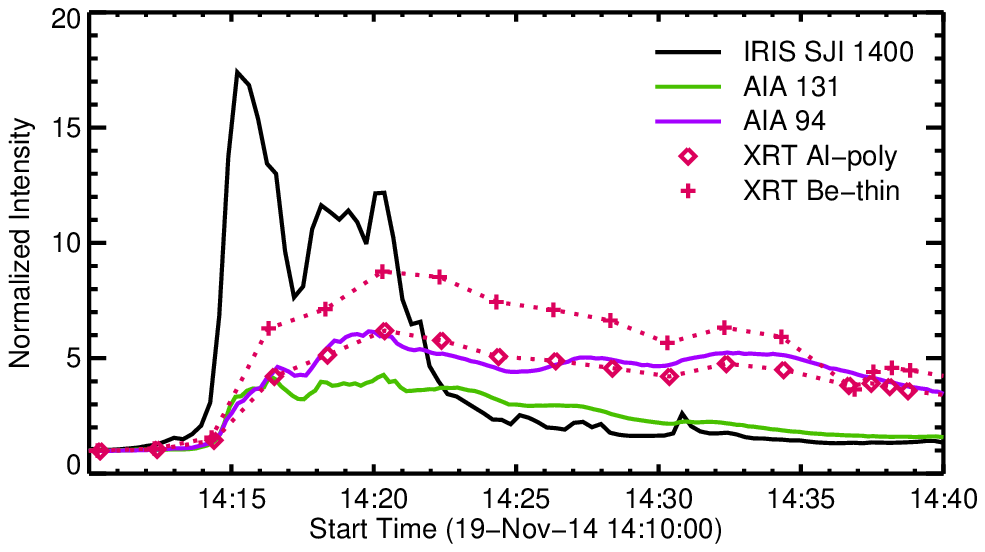}}
  \caption{(\textit{top panels}) The evolution of flare footpoints observed with \iris. The field
    of view shown is $26\arcsec\times26\arcsec$ in size. (\textit{bottom left panel}) The total
    intensity in the \iris\ slit jaw and in \rhessi\ as a function of time. Note that the
    \iris\ intensities are taken only from this small field of view while the \rhessi\ fluxes are
    not spatially resolved. Each light curve is normalized to its maximum. (\textit{bottom right
      panel}) The total intensity in \iris, \aia\ 94 \ion{Fe}{18}, \aia\ 131 \ion{Fe}{21},
    \xrt\ Al-poly/Open, and \xrt\ Be-thin/Open for the small field of view. Here each light curve
    is normalized to the intensity at the start of the observations.}
  \label{fig:rhessi}
\end{figure*}

The large volume of data returned by \iris\ makes it difficult to inspect every observation. To
filter the data we cross-referenced the flare catalog provided by the
\rhessiLong\ (\citealt{lin2002}, \rhessi) with the \iris\ observation catalog to find flares
for which there was \rhessi\ emission within the \iris\ slit-jaw field of view. We then created
quick-look movies of the \iris\ slit-jaw data and evaluated the events individually.

The microflare that occurred on 19 November 2014 beginning at about 14:14\,UT is particularly well
observed. In addition to data from \iris\ and \rhessi\ there are simultaneous observations from the
\eisLong\ (\eis, \citealt{culhane2007}), the \xrtLong\ (\xrt, \citealt{golub2007}), and the
\aiaLong\ (\aia, \citealt{lemen2012}). These data provide important constraints on the physical
properties in the loops that ultimately form above the footpoint regions observed with \iris\ and
\rhessi. The spectral lines observed with \eis, for example, provide information on both the
high-temperature loops through the observation of emission from \ion{Ca}{17}, \ion{Fe}{23}, and
\ion{Fe}{24} and the pressure in the loops through the \ion{Fe}{14} density diagnostic. These
observations are typically taken at relatively low cadence, with spectra taken at one position
every few minutes. Observations from \xrt\ and \aia\ in the 131\,\AA\ and 94\,\AA\ channels, which
observe \ion{Fe}{21} 128.75\,\AA\ and \ion{Fe}{18} 93.93\,\AA, provide information on the
high-temperature emission on much shorter time scales \citep[for a discussion of the \xrt\ and
  \aia\ temperature responses see][]{odwyer2014,odwyer2010}.

\begin{figure*}[t!]
  \centerline{\includegraphics[angle=90, width=\linewidth]{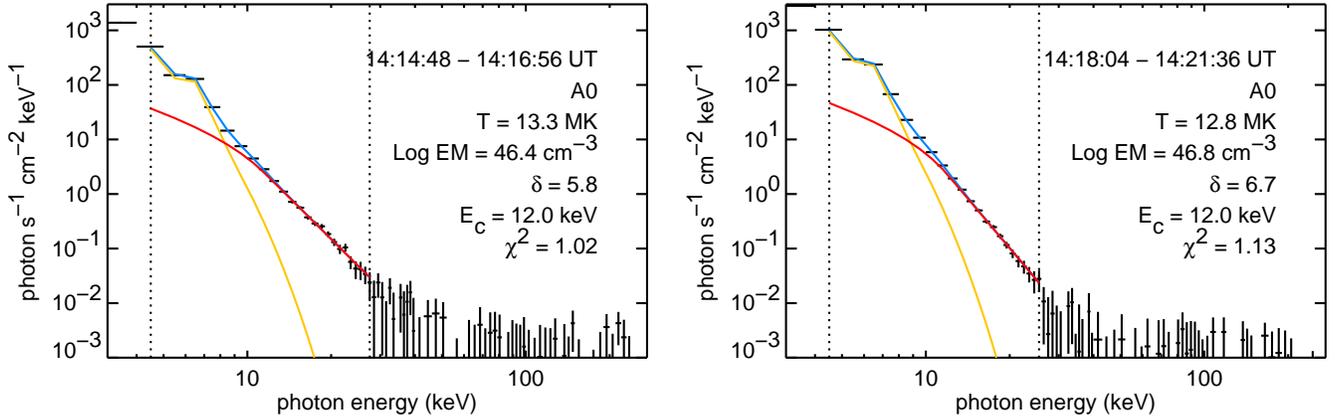}}
  \caption{RHESSI spectral fits of the two main peaks of the flare.}
  \label{fig:spectra}
\end{figure*}

\subsection{Observations of the Flare Footpoints}

In Figure~\ref{fig:context} we show an \aia\ 94\,\AA\ image, \hmi\ line-of-sight magnetogram, and
\iris\ 1400\,\AA\ slit-jaw image from the event, which occurred in the vicinity of a sunspot. The
\aia\ 94\,\AA\ image has been processed to remove some of the contaminating 1\,MK emission and
emphasize \ion{Fe}{18} \citep{warren2012}.  The larger \aia\ and \hmi\ images indicate the field of
view for the \iris\ slit-jaw image. For the period between 14:08 and 15:03\,UT \iris\ alternated
taking slit-jaw images in the 1400\,\AA\ and 1335\,\AA\ channels. The cadence for each channel was
about 19\,s.

The co-alignment of the various observations is an important component of this work. Unfortunately
the pointing information specified in most of the file headers is not accurate enough to
co-register these data. We have assumed that the \aia\ data have the most accurate pointing and
have written software to cross-correlate the \iris, \xrt\, and \eis\ observations to it. The
relatively high cadence of \aia\ --- the standard 12\,s for EUV images and 24\,s for UV images was
used during these observations --- insures that image pairs are always close together in time. For
the \iris\ slit-jaw images we co-align using \aia\ 1600\,\AA, for \xrt\ we use \aia\ 94\,\AA, and
for \eis\ 195.119\,\AA\ we use \aia\ 195\,\AA. For the lower resolution \xrt\ and \eis\ data,
blinking the images indicates that this procedure works very well. For the \iris\ data it is able
to correct for longer-term pointing drifts in the image sequence but does introduce some jitter
that is evident in the animations of the data. We note that no adjustments to the \rhessi\ pointing
appear to be necessary.

The \iris\ slit-jaw images for this period show intense brightenings in the sunspot penumbra and in
some of the near-by opposite polarity flux. To investigate the relationship between these
brightenings and the hard X-ray emission we have computed a \rhessi\ 15--25\,keV image using the
``clean'' algorithm \citep{hurford2002} with a 2.4\arcsec\ pixel size and a 240\,s integration
centered on the time of each \iris\ slit-jaw image.  The long integration time relative to the
cadence of the \iris\ slit-jaw observations is necessary to bring up the signal-to-noise.  Only
detectors 3, 6, 8, and 9 are used for these image reconstructions. Except for the
\verb+clean_beam_width_factor+, which we set to 1.5 to narrow the spatial extent of the beam, we
use the default \rhessi\ imaging parameters for the clean algorithm.

\begin{figure}[t!]
  \centerline{\includegraphics[width=0.5\linewidth]{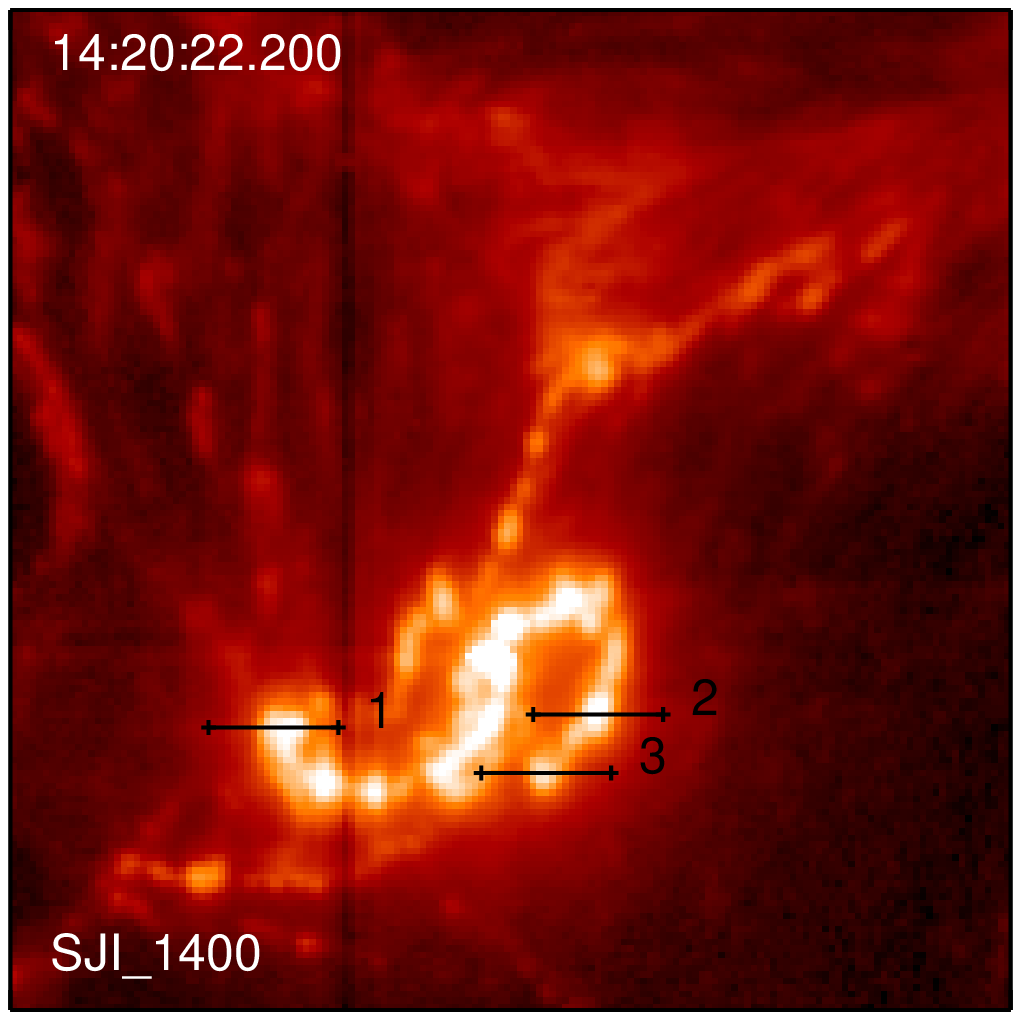}%
    \includegraphics[width=0.5\linewidth]{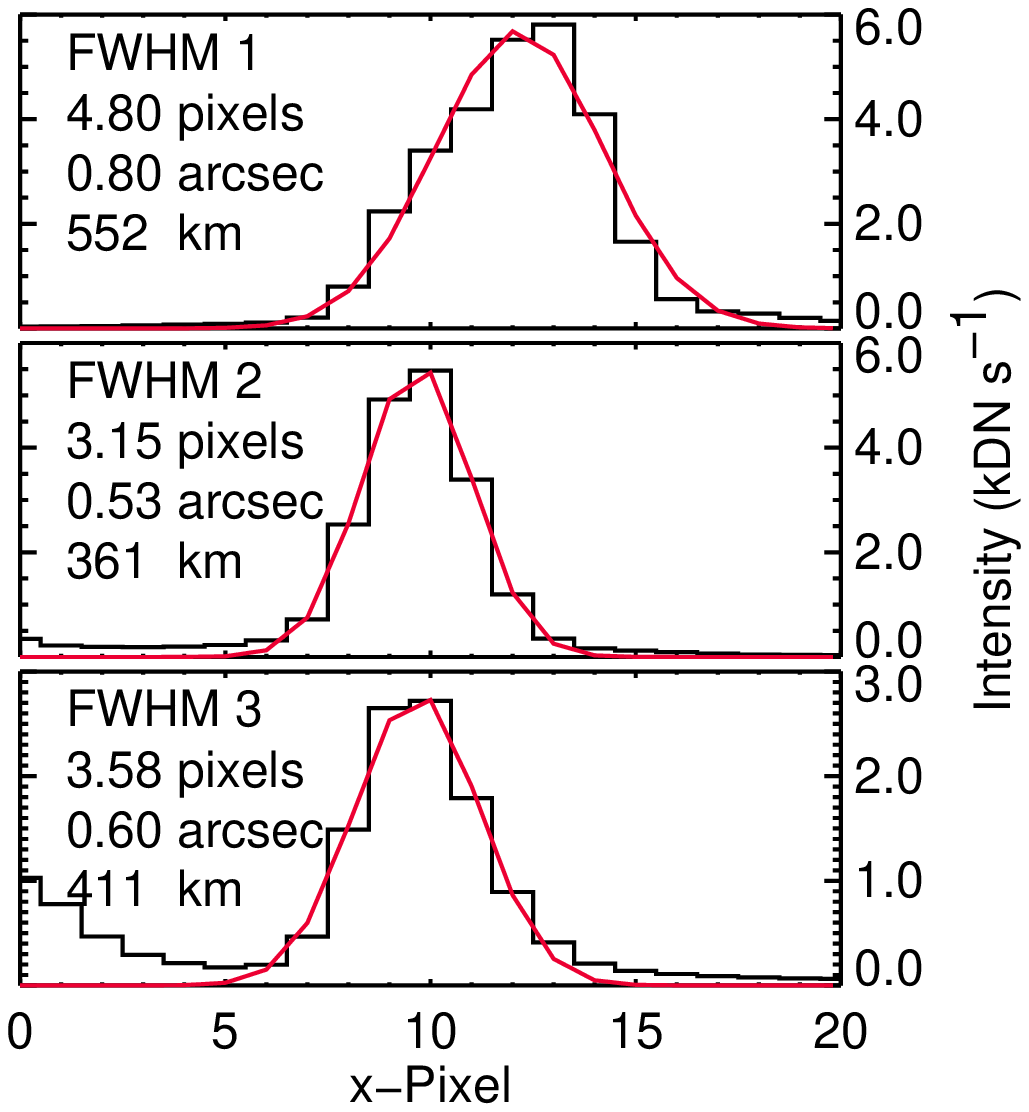}}
  \vspace{0.2in}
  \caption{Cross-sectional profiles of IRIS footpoint intensities at 14:20 UT. The left panel shows
     cross-sectional lines through three footpoint brightenings. The right panel shows the
     corresponding intensity profiles and FWHM.}
  \label{fig:fwhm}
\end{figure}

\begin{figure}[t!]
  \centerline{\includegraphics[width=\linewidth]{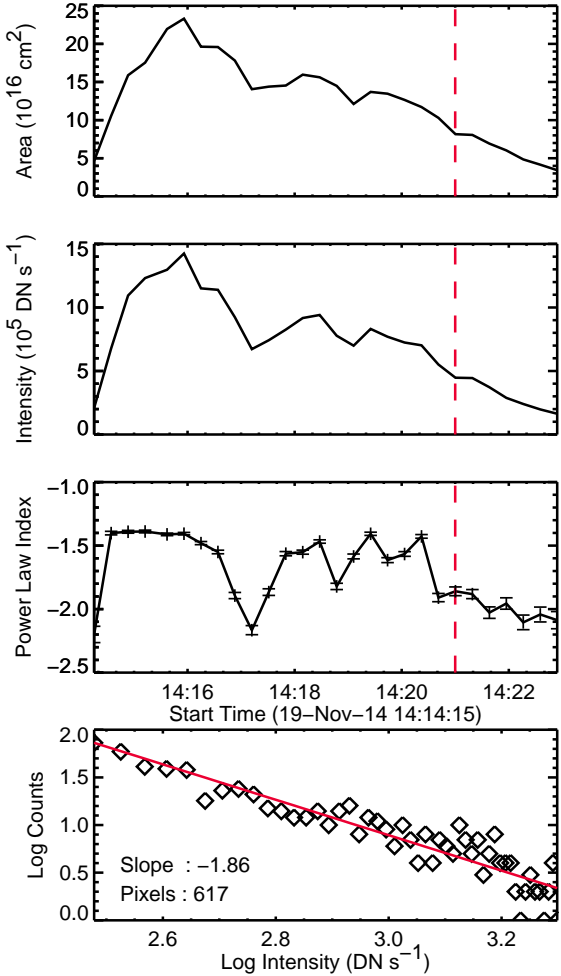}}
  \caption{The distribution of \iris\ footpoint intensities at 14:21\,UT (bottom panel). The top
    panels show the total footpoint area, the total footpoint intensity, and the power-law index of
    the footpoint intensity as functions of time. These quantities are derived from the footpoints
    near the sunspot penumbra.}
  \label{fig:hist}
\end{figure}

\begin{figure*}[t!]
  \centerline{\includegraphics[angle=90,width=\linewidth]{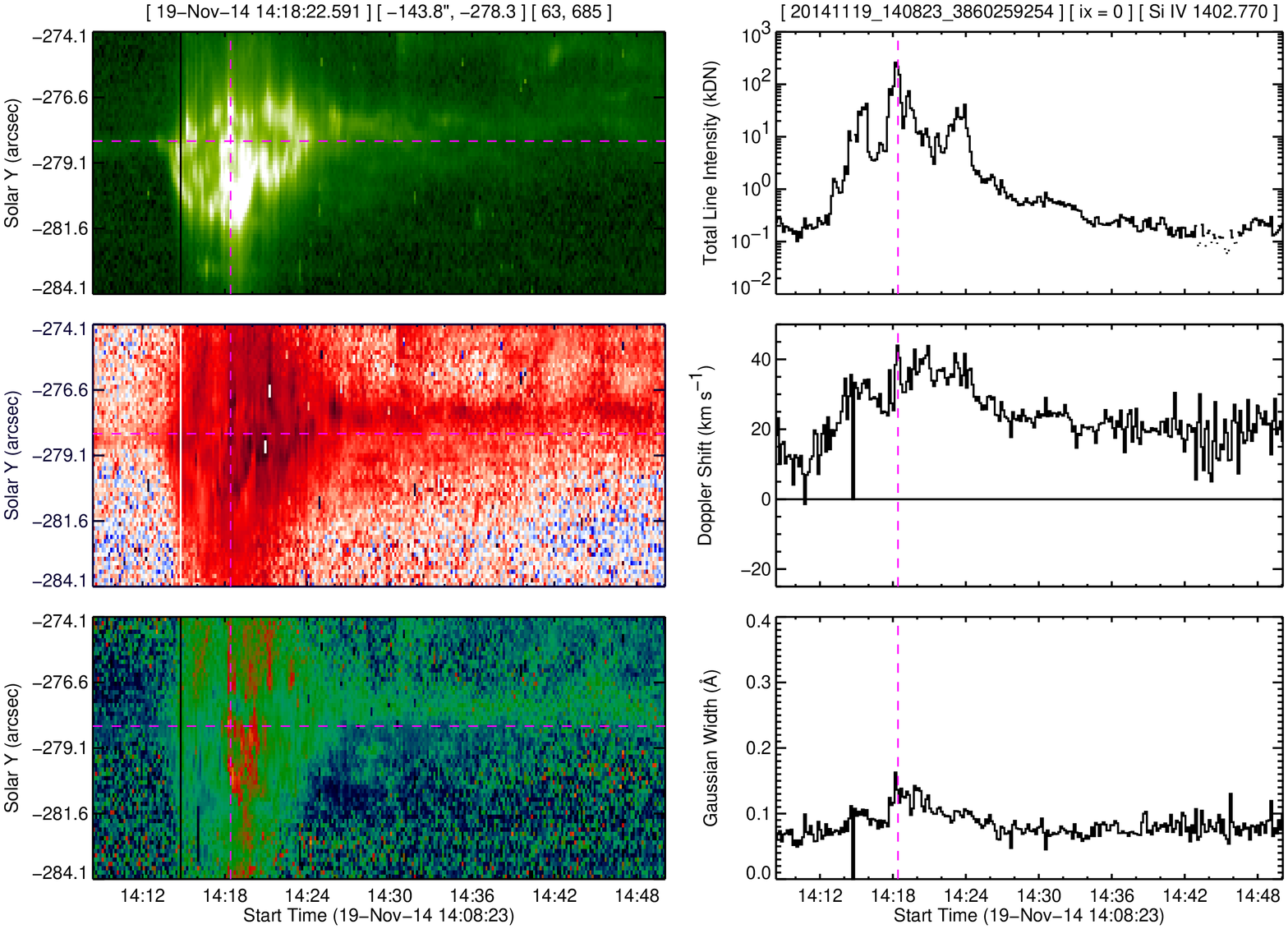}}
  \caption{Intensities, Doppler shifts, and Gaussian line widths determined from moments of the
    \ion{Si}{4} 1402.770\,\AA\ line. The line profiles are red-shifted at almost all times from
    about 14:14 to 14:24, and some regions show red-shifts to 14:50. The left panels show these
    quantities as a function of position along the slit and time. The region along the slit
    corresponds to the brightenings seen near the sun spot penumbra. Note that the values listed in
    brackets above these panels indicate the position of the horizontal line in both absolute
    coordinates co-aligned to AIA (-143.8\arcsec, -278.3\arcsec) and in the pixel coordinates of
    the data array (63, 685). The right panels show these quantities for a single slit position as
    a function of time. The vertical line indicates 2014 November 19 14:18:22.591\,UT.}
  \label{fig:si4}
\end{figure*}

The \rhessi\ images indicate that most of the hard X-ray emission comes from the footpoints rooted
in the strong magnetic field of the sunspot penumbra. Figure~\ref{fig:context} shows the contours
of the \rhessi\ emission superimposed on an \iris\ slit-jaw image. The lowest intensity
\rhessi\ contour shown is about a factor of two smaller than the peak. Lower intensity values show
features that are clearly unrelated to the structures seen in the \iris\ images and likely to be
noise generated from the inversion.

In Figure~\ref{fig:rhessi} we show individual \iris\ slit-jaw images as well as light curves
computed by summing the \iris\ intensities over the small region of interest and integrating all of
the available \rhessi\ counts. As expected, these light curves show a close relationship between
the transition region intensity and the hard X-ray emission. The images and light curves show that
hard X-ray emission is concentrated into two bursts that peak at approximately 14:15 and
14:20\,UT. Isothermal fits to the \rhessi\ spectra for times centered around these peaks are shown
in Figure~\ref{fig:spectra}. Here the \verb!vth+thick2_vnorm!  model is used. This model describes
an isothermal component plus a non-thermal component produced by thick target bremsstrahlung from a
power-law distribution of electrons. It is part of the standard spectral analysis software for
\rhessi\ \citep{schwartz2002}.

The best-fit temperatures, which will be relevant to the discussion of the temperatures derived
from the \xrt, \eis, and \aia\ observations later in the paper, are $13.3\pm0.4$ and
$12.8\pm0.3$\,MK. The spectral indexes for the non-thermal component of the electrons for these
times are $5.8\pm0.1$ and $6.7\pm0.2$, and are important parameters in modeling the energy
deposition in these footpoint regions.

The differences in spatial resolution between the two instruments are readily apparent here. The
broad regions of hard X-ray emission are imaged as many small footpoint brightenings in \iris. As
shown in Figure~\ref{fig:fwhm}, the footpoints observed in \iris\ typically have a FWHM of about
$0\farcs6$ or about 410\,km. Equally important is the difference in dynamic range. The
\iris\ detector records data values to 14 bits (0 to 16,383 or about 4 orders of magnitude) while
\rhessi\ has a dynamic range of about a factor of two in this event.

The strong correlation between the transition region and hard X-ray emission suggests that we can
gain additional insight into the distribution of heating events along footpoints by measuring the
distribution of \iris\ intensities there.  We measured this distribution for the duration of the
hard X-ray burst, finding it well described by a power-law and well correlated with the hard X-ray
intensity.  The intensity histogram for an \iris\ image taken near the second hard X-ray peak shown
in Figure~\ref{fig:hist} illustrates the power-law distributions that are observed in the
footpoints. The indexes on these distributions range between -1.5 and -2.5 and have a median value
of about -1.6. Note that this index is fundamentally different than the spectral index on the
non-thermal electron distribution. This index describes how the energy released during the flare is
distributed across different field lines while the spectral index describes how energy is
distributed across all of the electrons.

\begin{figure*}[t!]
  \centerline{\includegraphics[angle=90,width=\linewidth]{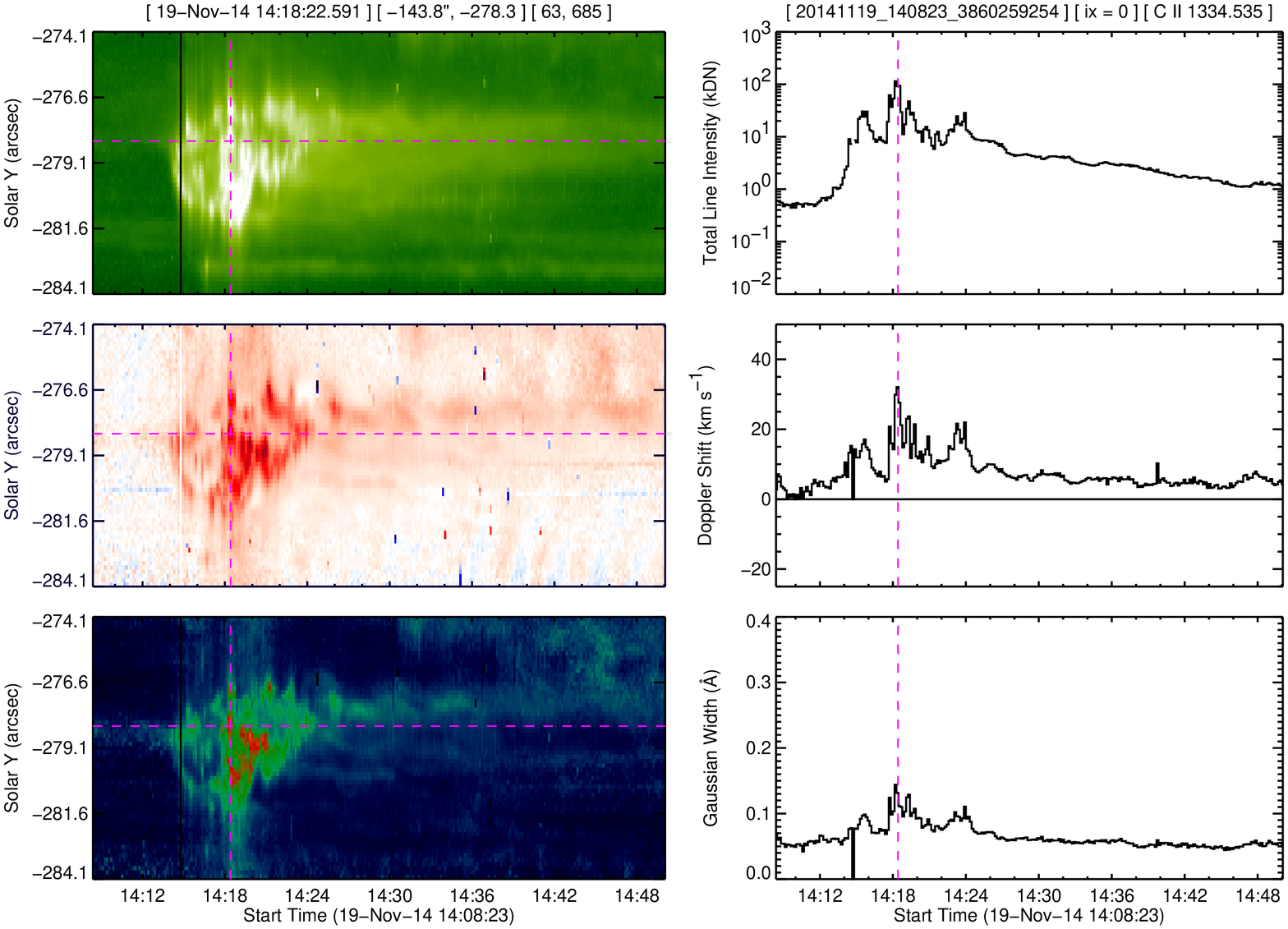}}
  \caption{The same format as Figure~\ref{fig:si4} but for the \ion{C}{2} 1334.535\,\AA\ line. The
    variations in intensity, Doppler shift, and line width are well correlated with those measured
    for \ion{Si}{4}, but the magnitude of the response is somewhat smaller.}
  \label{fig:c2}
\end{figure*}

\begin{figure*}[t!]
  \centerline{\includegraphics[angle=90,width=\linewidth]{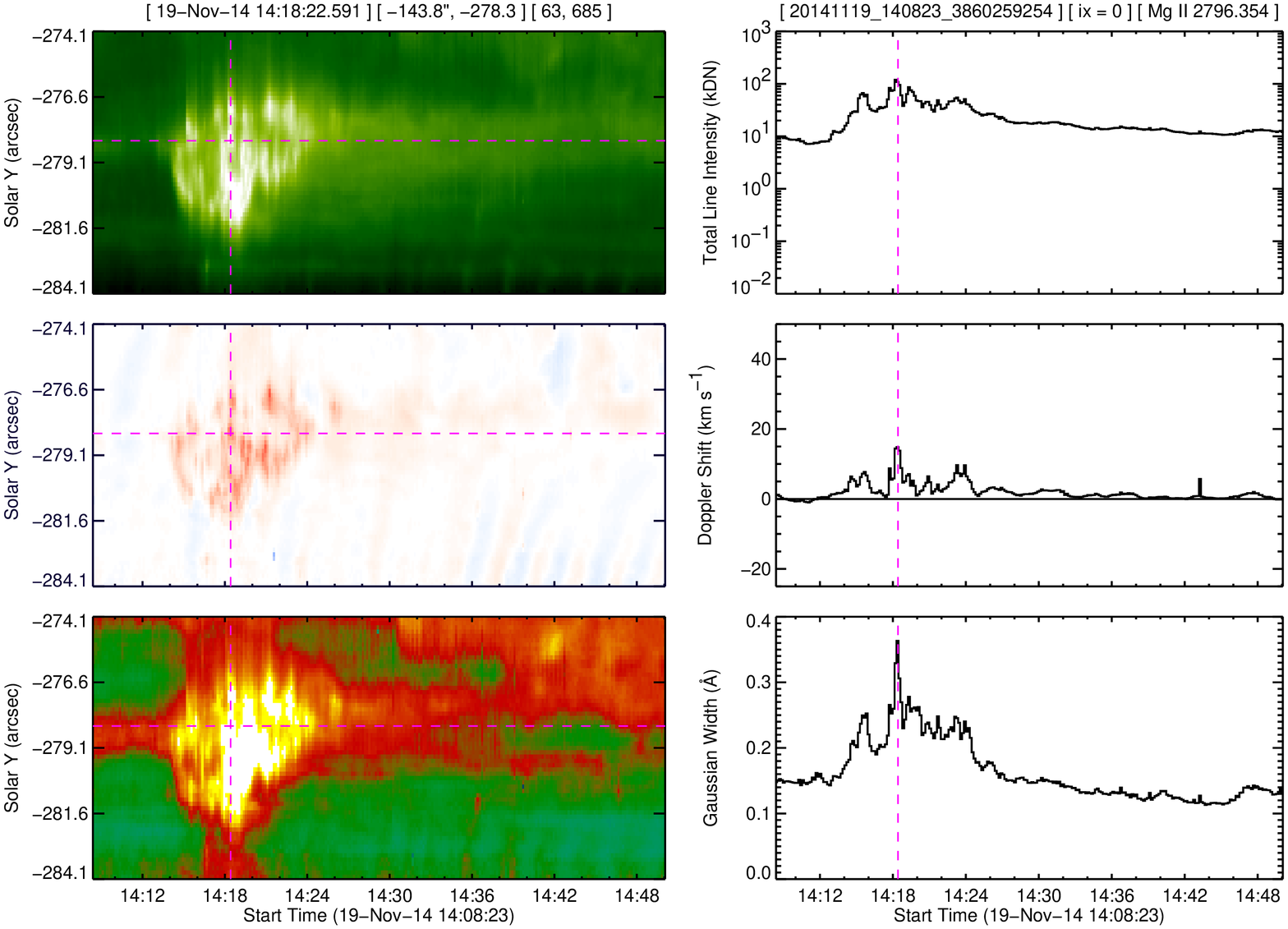}}
  \caption{The same format as Figure~\ref{fig:si4} but for the \ion{Mg}{2} 2796.354\,\AA\ line. The
    variations in intensity, Doppler shift, and line width are well correlated with those measured
    for \ion{Si}{4}, but the magnitude of the response is somewhat smaller.}
  \label{fig:mg2}
\end{figure*}

\begin{figure*}[t!]
  \centerline{\includegraphics[angle=90,width=0.9\linewidth]{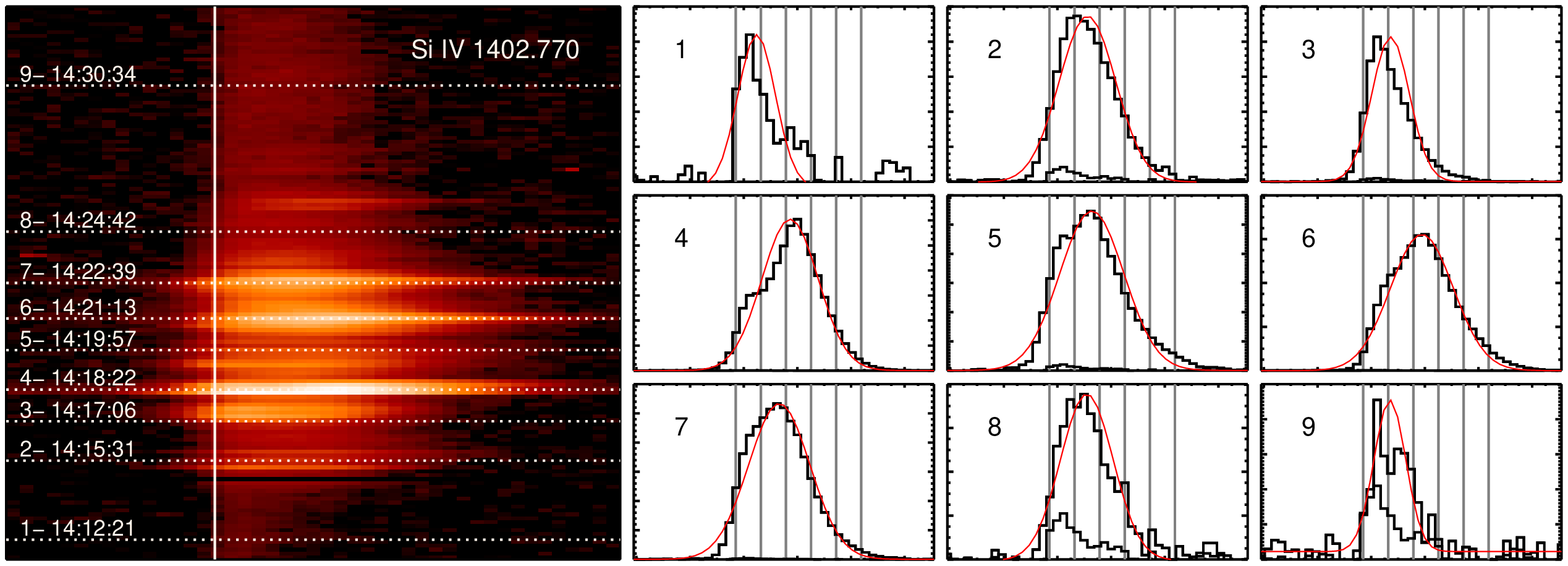}}
  \centerline{\includegraphics[angle=90,width=0.9\linewidth]{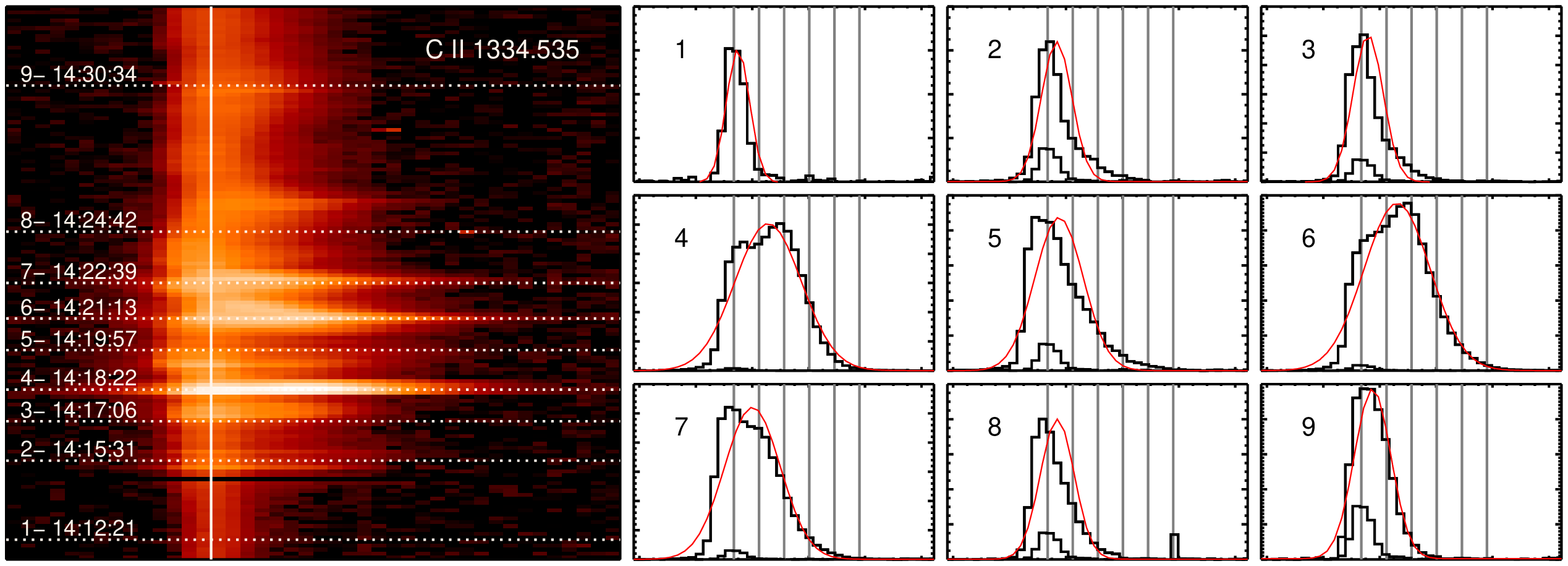}}
  \centerline{\includegraphics[angle=90,width=0.9\linewidth]{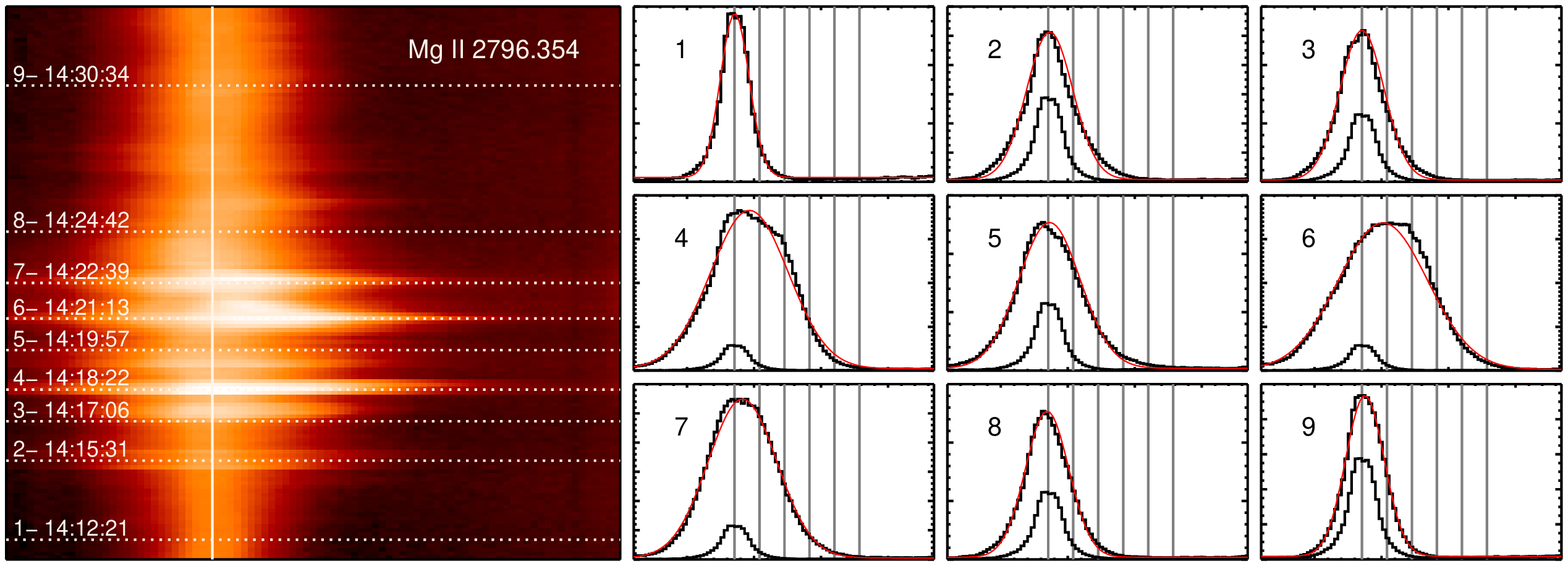}}  
  \caption{IRIS spectra as a function of time for \ion{Si}{4} 1402.770\,\AA, \ion{C}{2}
    1334.535\,\AA, and \ion{Mg}{2} 2796.354\,\AA\ for a single position. This figure illustrates
    the multi-component nature of the profiles observed during the flare. The left panels show a
    stackplot of the spectra while the panels on the right show profiles for selected times. The
    vertical lines on the spectral profiles are at velocities of 0, 20, 40, 60, 80, and
    100\,km~s$^{-1}$. Also shown in each profile is the Gaussian profile inferred from the moment
    calculation. To give a sense of the intensity at each time, a pre-flare profile (\#1) is also
    shown in panels 2--9. The $y$ position for these spectra is indicated by the horizontal line in
    Figures~\ref{fig:si4}, \ref{fig:c2}, and \ref{fig:mg2}.}
  \label{fig:prof}
\end{figure*}

The \iris\ slit provides another view on the physical conditions in the footpoints. During this
period the slit was kept in a fixed position and spectra were recorded at a cadence of about
9.5\,s. To make the data easier to visualize we computed moments for each line profile. The moments
are defined as
\begin{eqnarray}
  I & = & \sum_{i=1}^{N} I_i \\
  \lambda_0 & = & \frac{1}{I} \sum_{i=1}^{N} \lambda_i I_i \\
  \sigma & = & \frac{I\Delta}{\textrm{max}(I_i)\sqrt{2\pi}},
\end{eqnarray}
where $I_i$ is the background subtracted intensity in units of DN s$^{-1}$ and $\Delta$ is the
width of a spectral pixel. This alternative definition of the width moment is used because it
yields better performance for low count rates. We chose to compute moments instead of fitting the
profiles because the moment calculation is much faster and, as we will see, the profiles are not
particularly well represented by a single Gaussian. We could fit multiple Gaussians to these
profiles, but such fits are often poorly constrained.

We focus on the moment calculations for \ion{Si}{4} 1402.770\,\AA, \ion{C}{2} 1334.535\,\AA, and
\ion{O}{1} 1355.598\,\AA. These lines are formed at progressively lower temperatures. In ionization
equilibrium, \ion{Si}{4} peaks at $\log T=4.8$, \ion{C}{2} at 4.4, and \ion{O}{1} at or below
4.0. We also consider the spectral region near the \ion{Mg}{2} h \& k doublet, which is sensitive
to a wide range of temperatures in the solar chromosphere \citep[e.g.,][]{pereira2013}. These
spectral features are optically thick and often show a strong reversal in the core of the line
\citep[e.g.,][]{schmit2015}. As we will see, the profiles are not reversed during this flare and
the Gaussian moment is an adequate description of the line profile. We take the rest wavelengths
for the doublet, 2796.354\,\AA\ for the k-line and 2803.531\,\AA\ for the h-line, from
\citet{murphy2014}. Fitting time-averaged profiles from the sunspot in this observation yields
velocities of less than 1\,km s$^{-1}$ with these wavelengths.

We also investigated \ion{Fe}{12} 1349.382\,\AA\ (6.2) and \ion{Fe}{21} 1354.080\,\AA\ (7.05), but
did not detect emission in these lines. We will discuss the coronal component of these observations
in section \ref{sec:hot}.

In Figure~\ref{fig:si4}, \ref{fig:c2}, and \ref{fig:mg2} we show the intensity, Doppler velocity,
and width as a function of space and time for the \ion{Si}{4}, \ion{C}{2}, and \ion{Mg}{2}
lines. The results for \ion{O}{1} are not shown. These data are taken from the region around the
southern brightening, which is the only flare-related emission observed along the slit.

During the event the intensity in \ion{Si}{4} rises by about a factor of 1000, the intensity in
\ion{C}{2} rises by about a factor of 100, and the \ion{Mg}{2} and \ion{O}{1} intensity rises by a
factor of 10. During the initial part of the event, from approximately 14:14 to 14:24\,UT, numerous
increases in the intensity lasting about 30\,s are observed. These intensity bursts are well
correlated among the various emission lines. After about 14:24 UT there is a slow decay phase
during which the fluctuation level in the intensity of all of the lines is much lower.

During the event we also observe systematic red shifts in \ion{Si}{4} and \ion{C}{2}. In
\ion{Si}{4} the red shifts are typically 10--40\,km~s$^{-1}$. In \ion{C}{2} the red-shifts are
somewhat smaller, at 5--30\,km~s$^{-1}$. \ion{Mg}{2} shows red shifts associated with the strongest
bursts. The \ion{O}{1} line shows essentially no change in velocity during the flare. This line
typically shows a blue shift of about 2\,km~s$^{-1}$ at all times, which likely reflects the
uncertainty in the wavelength calibration.

The red-shifts observed in \ion{Si}{4} and \ion{C}{2} persist at elevated levels throughout the
early part of the event. From approximately 14:14 to 14:24\,UT almost all of the intense emission
in these lines is red-shifted. After this time, the persistent red-shifts appear to be more
localized. There is a region around $y\sim-277\arcsec$ where the red-shifts last beyond 14:50 UT
in both lines.

For \ion{Si}{4} and \ion{C}{2} we also observe changes in the line width during the event. These
changes are similar in magnitude for both lines and well correlated with the intensity
fluctuations. In contrast with the Doppler shift, the line width observed for Si IV and C II is
close to the pre-flare value during the decay of the event. Again, \ion{O}{1} behaves differently,
showing essentially no change in width during the flare.

\begin{figure*}[t!]
  \centerline{%
    \includegraphics[width=0.95\linewidth]{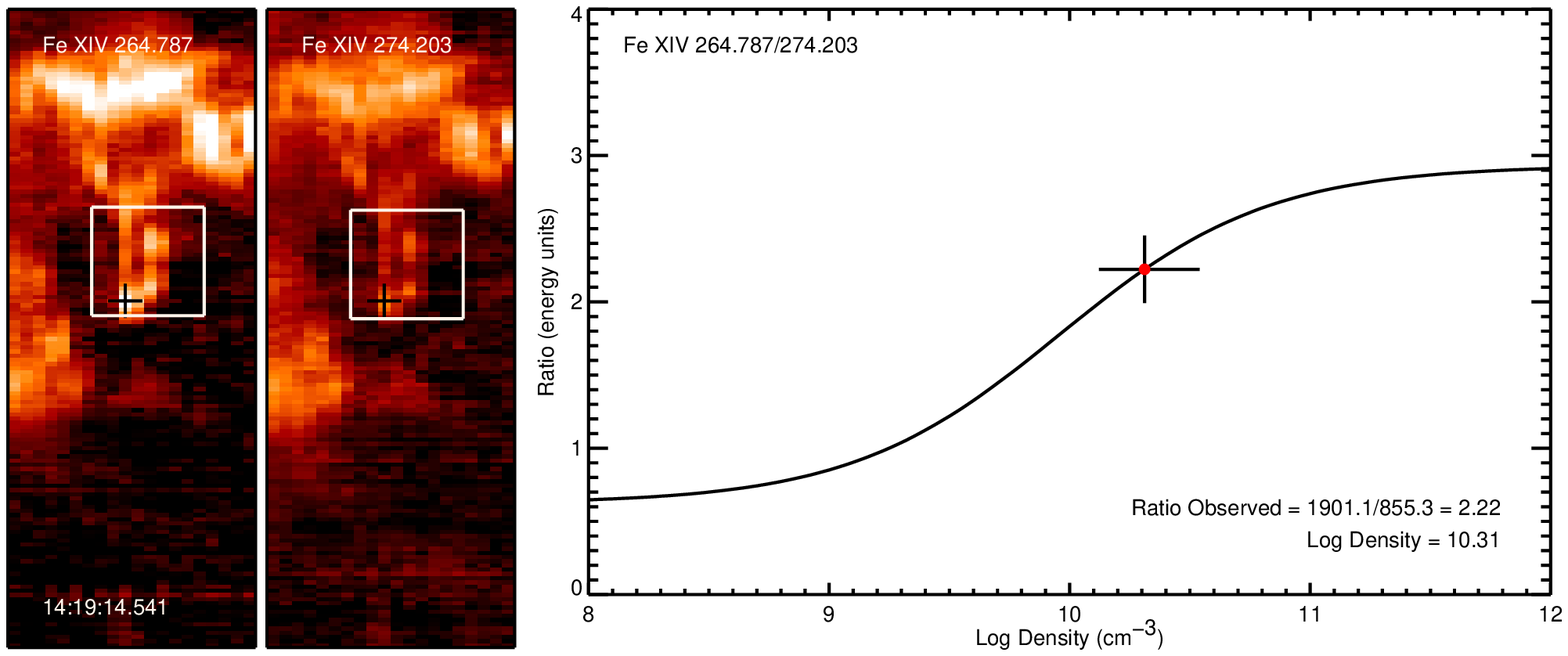}}
  \caption{The electron density in the footpoint region derived from the EIS \ion{Fe}{14} density
    diagnostic. The intensities were measured in an 9\,s exposure that began at 14:19:14\,UT and
    indicate a pressure of about $\log P = 16.9$.  The intensities have units of erg cm$^{-2}$
    s$^{-1}$ sr$^{-1}$.}
  \label{fig:eis}
\end{figure*}

The moments give us a sense of how the bulk properties of the line profiles are evolving during the
flare, but they also act to compress the data and may obscure important spectral features. We have
not, for example, attempted to account for any unrelated emission along the line of sight and the
observed profile could be a mixture of flare and non-flare components.

In Figure~\ref{fig:prof} we display \ion{Si}{4}, \ion{C}{2}, and \ion{Mg}{2} line profiles as a
function of time at a single spatial location. Here we see the line profiles appear to be composed
of two or more distinct components. We also see that the emission in these components is
significantly enhanced over the background at all wavelengths during the flare. For \ion{Si}{4},
one of the components typically has a velocity of about 20\,km~s$^{-1}$ while the other is at about
40\,km~s$^{-1}$. \ion{C}{2} also exhibits multiple components, but one is typically at rest while
the other primary component is typically at about 20--40\,km~s$^{-1}$. The \ion{Mg}{2} profiles are
similar to those of \ion{C}{2}. As is suggested by the lack of variation in the width, \ion{O}{1}
does not show any complex dynamical evolution in the line profile.

The final piece of information on the physical conditions in the footpoints that we have considered
comes from the \eis\ electron densities inferred from the \ion{Fe}{14}
264.787/274.203\,\AA\ ratio. As is indicated by the theoretical ratio taken from version 8 of
CHIANTI \citep{delzanna2015}, this line ratio is sensitive to density up to about $\log n_e =
11$. Since this ion is formed at about 2\,MK, this ratio can measure pressures to about $\log
P = 2n_eT_e = 17.6$, where the pressure has units of cm$^{-3}$~K. The \eis\ observations for this
period consisted of a sparse scan across the active region using 3\arcsec\ steps. The spatial
resolution along the slit was the nominal 1\arcsec\ per pixel. At each of the 20 positions a 9\,s
exposure was taken. The resulting cadence for the scan, which included the detector readout, was
about 3\,min 30\,s. We fit each profile in the rasters taken between 14:03 and 14:53 with
Gaussians. As is shown in Figure~\ref{fig:eis}, the highest densities during this time were
observed during an exposure taken at 14:19:14\,UT and indicate a pressure in the loop of $\log P =
16.9$. The pressure is an important constraint because it is directly related to the rate at which
energy is deposited in the loop. The absolute intensity, in contrast, depends on both the heating
rate and the filling factor \cite[e.g.,][]{warren2008}.

\subsection{Observations of the Flare Loops}\label{sec:hot}

\begin{figure*}[t!]
  \centerline{\includegraphics[width=0.95\linewidth]{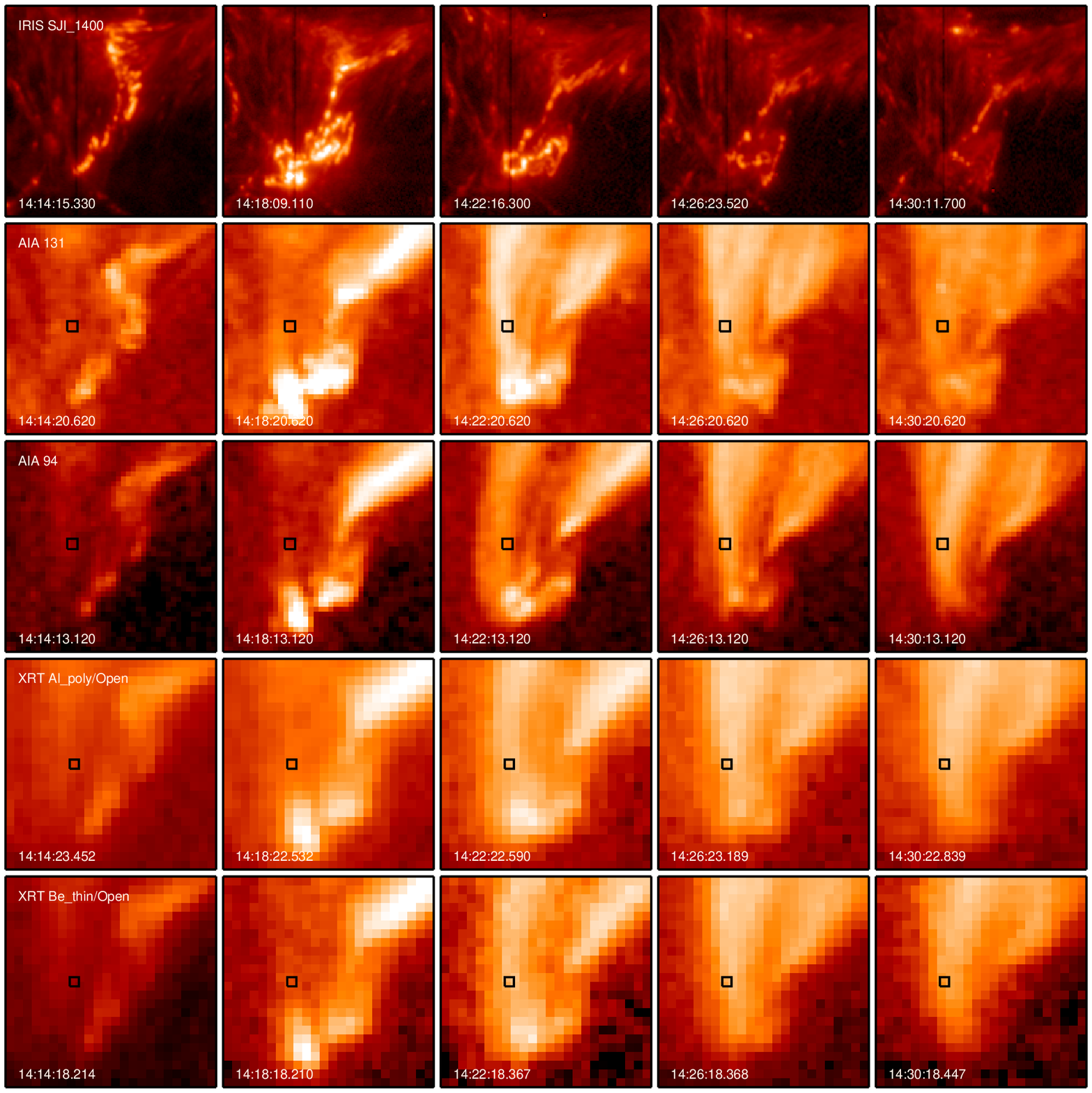}}
  \caption{Observations of the region around the southern flare footpoint with \iris, \aia, and
    \xrt. The field of view is $26\arcsec\times26\arcsec$ in size.}
  \label{fig:smallfov}
\end{figure*}

\begin{figure*}[t!]
  \centerline{%
    \includegraphics[width=0.5\linewidth]{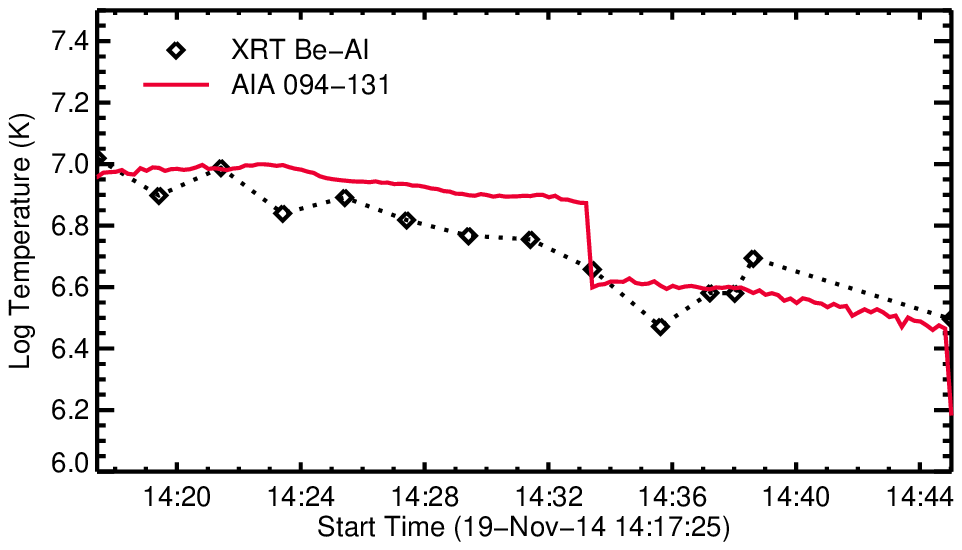}%
    \includegraphics[width=0.5\linewidth]{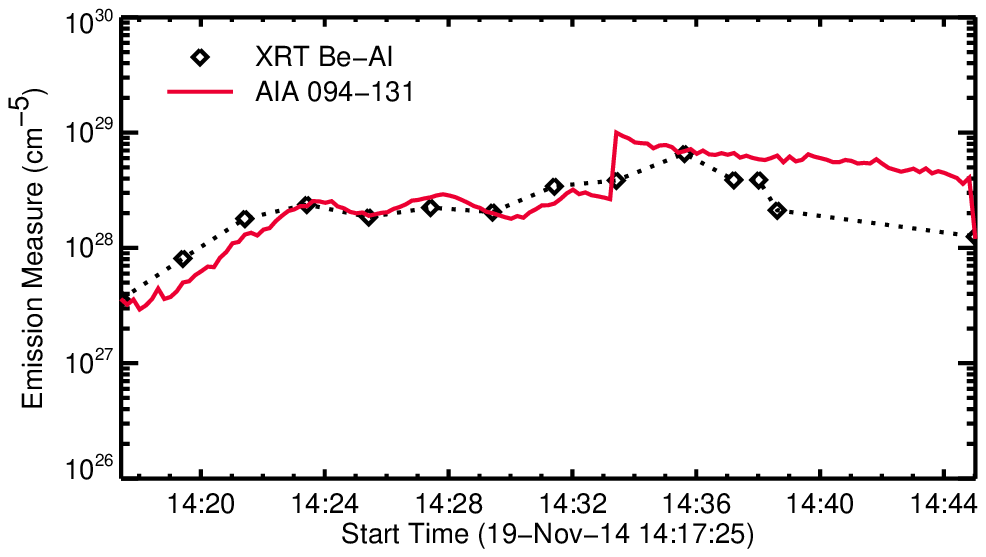}}
  \caption{\xrt\ and \aia\ temperatures and emission measures as a function of time derived from
    \xrt\ BeThin/AlPoly and \aia\ 131/94 filter ratios. The \aia\ ratio can be multi-valued and we
    chose the temperature closest to that derived from the closest \xrt\ measurement. This is the
    origin of the discontinuity near 14:33.}
  \label{fig:teem}
\end{figure*}

\begin{figure*}[t!]
  \centerline{%
    \includegraphics[width=0.5\linewidth]{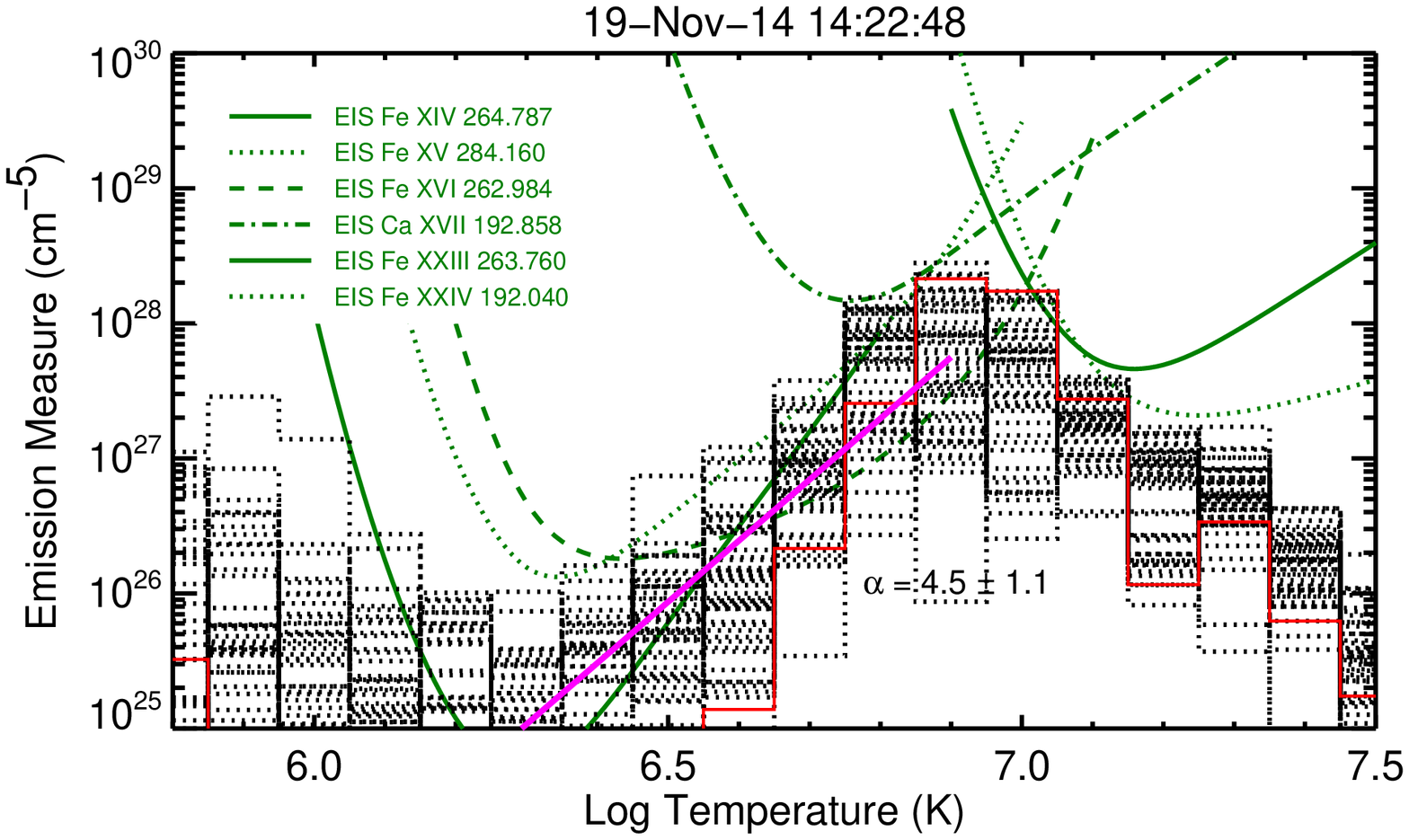}
    \includegraphics[width=0.5\linewidth]{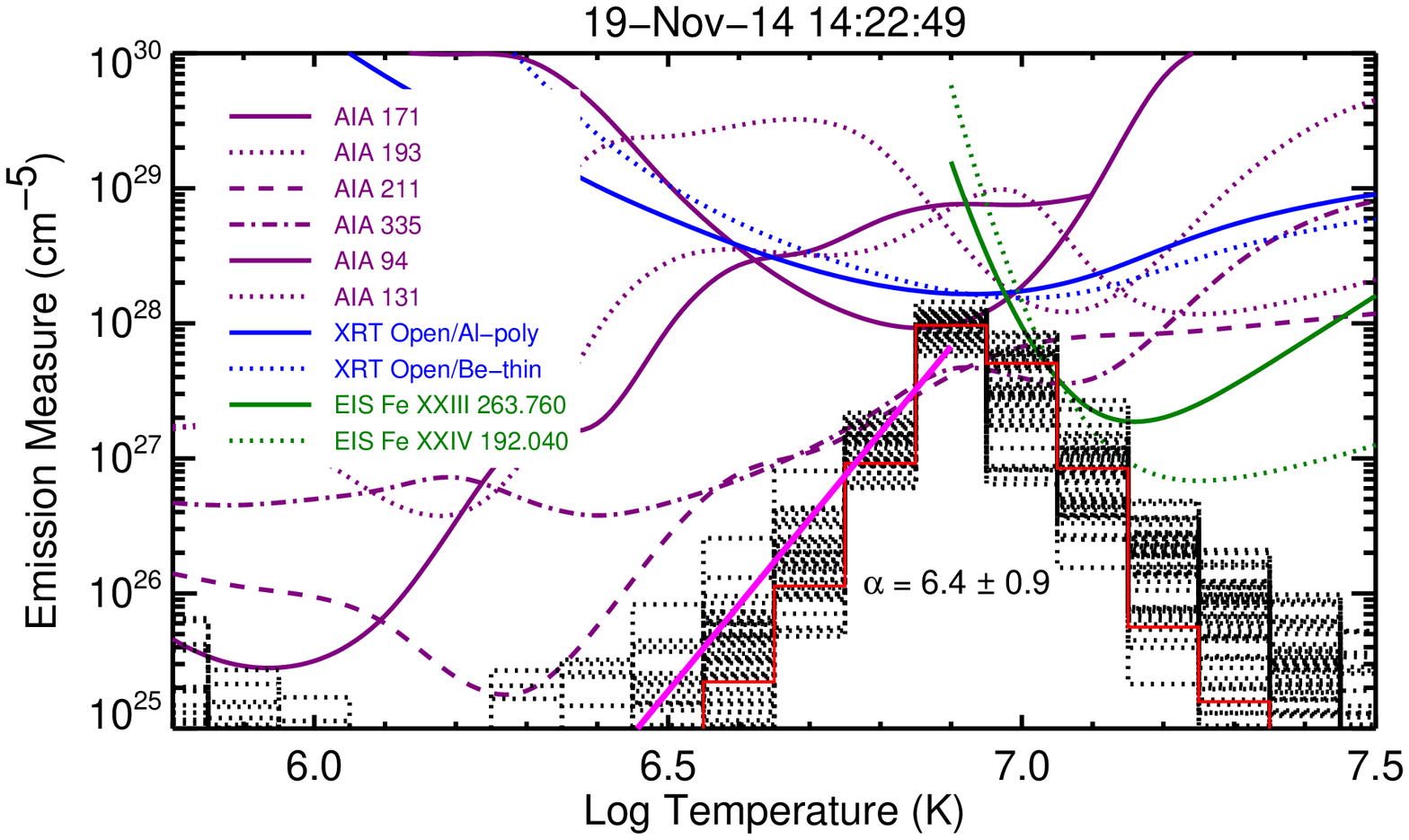}}
  \centerline{%
    \includegraphics[width=0.5\linewidth]{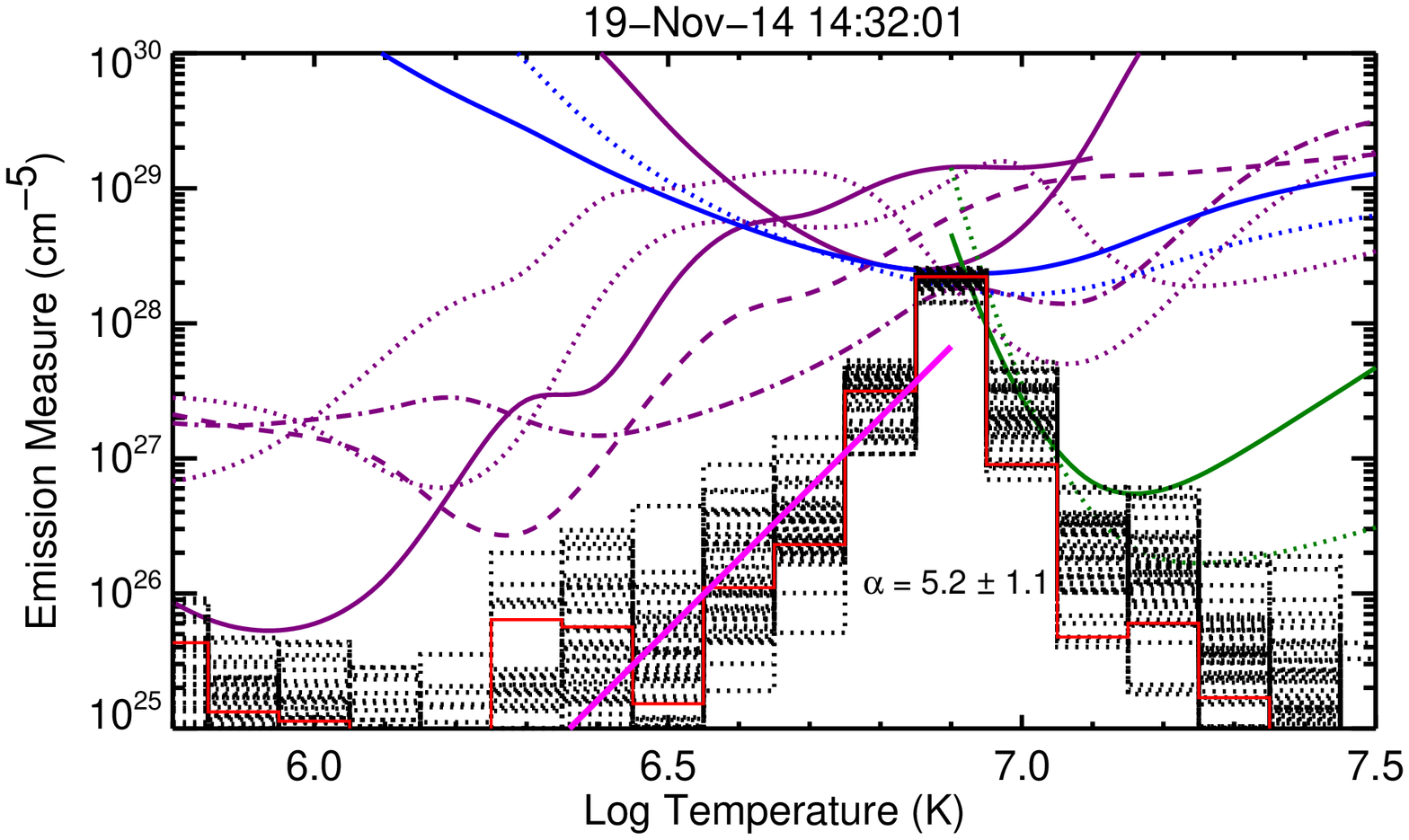}
    \includegraphics[width=0.5\linewidth]{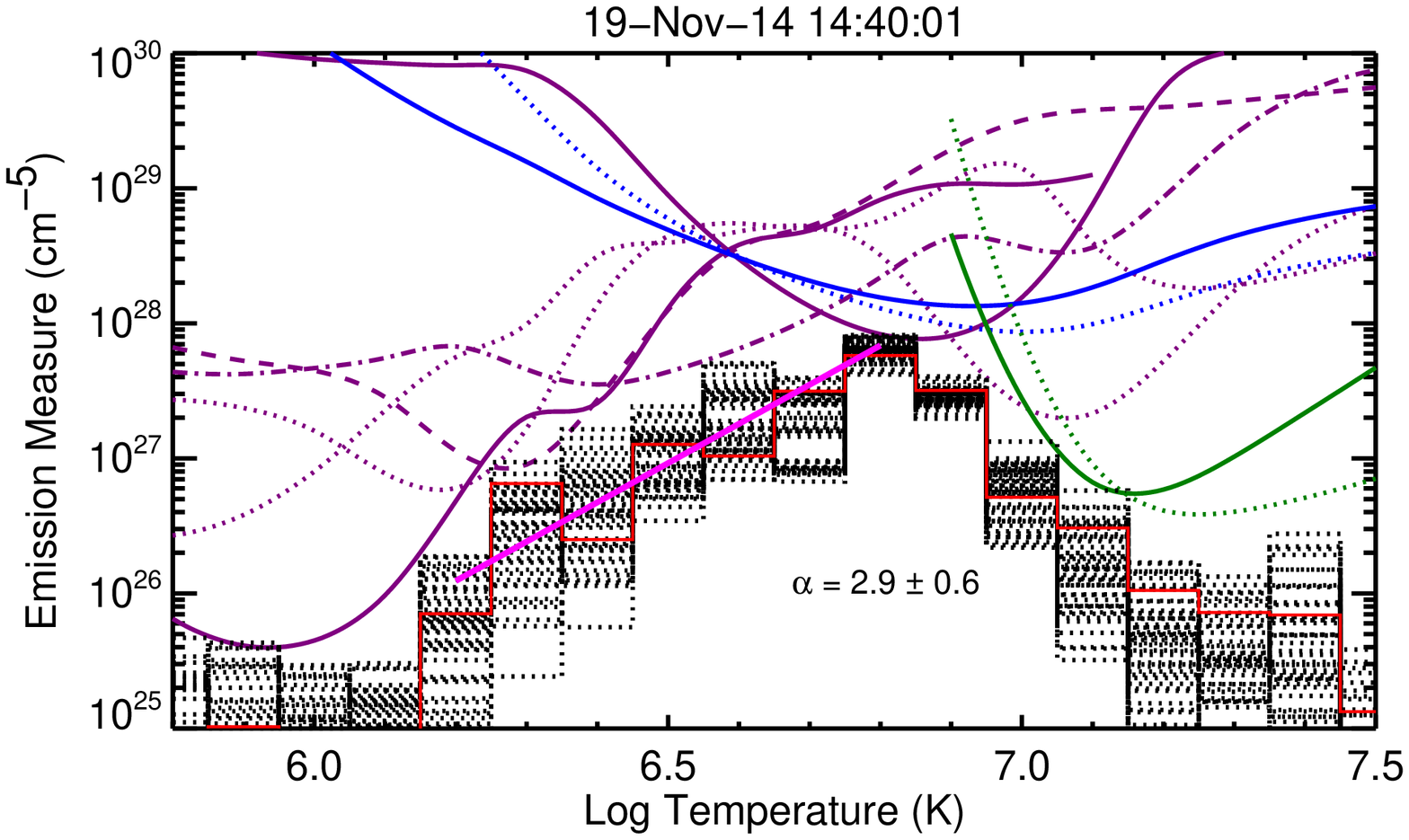}}
  \caption{The differential emission measure distribution computed from \eis, \aia, and \xrt\ for
    several times during the event. The distributions generally peak at about 10\,MK and fall off
    sharply at both higher and lower temperatures.}
  \label{fig:dem}
\end{figure*}

High-temperature flare loops are rooted in the footpoint regions observed with \iris, \rhessi, and
\eis. We can determine the properties of these loops directly through the observations available
from \xrt, \aia, and \eis. During this time \xrt\ took full-resolution images in the Al-Poly/Open
and Be-Thin/Open filter combinations at a cadence of about 120\,s. \xrt\ has a plate scale of about
1\arcsec\ per spatial pixel. As noted earlier, \aia\ took full-resolution images in all of the EUV
channels at a cadence of about 12\,s and \eis\ executed a sparse raster over the active region. In
addition to the \ion{Fe}{14} lines discussed previously, \eis\ recorded the \ion{Fe}{16} 262.984,
\ion{Fe}{23} 263.760, and \ion{Fe}{24} 192.040\,\AA.

In Figure~\ref{fig:smallfov} we show selected images from \xrt\ and \aia\ from the small footpoint
region near the sunspot penumbra. The temporal evolution of the intensities from this area is shown
in Figure~\ref{fig:rhessi}. As one would expect, the high temperature emission peaks
later than the footpoint emission observed with \iris\ and \rhessi\ \citep{neupert1968}.

The key constraint that we hope to derive from these observations is a quantitative measure of the
temperature distribution for the loops connected to the IRIS footpoints. We measure the temperature
in two ways, first we compute filter ratios and then we consider the differential
emission measure distribution.

For these calculations we consider the XRT, AIA, and EIS intensities in a small region just above
the footpoints imaged in the IRIS slit. This is indicated by the black box shown in
Figure~\ref{fig:smallfov}. Ideally we would consider a point at the loop apex, but further away
from the footpoints it becomes unclear if all of the high-temperature emission we observe connects
to this footpoint region (see the AIA 94\,\AA\ image in Figure~\ref{fig:context}). We expect the
temperature gradients in the corona to be relatively flat and it is unlikely that there will be
significant differences between the region we have chosen and the actual loop apex. We estimate the
length of these loops to be about 40\,Mm.

For the filter ratios we consider \xrt\ BeThin/AlPoly and \aia\ 131/94. As shown in
Figure~\ref{fig:teem}, the temperatures derived from these two ratios are generally consistent. The
temperatures begin at about $\log T = 7.0$ and decline throughout the event. The \aia\ filter
ratios can be multi-valued and we have chosen the temperature closest to that inferred from
\xrt. The emission measures derived from these ratios have very similar trends but differ in
magnitude by about a factor of 2. We have divided the \xrt\ intensities by 2 to bring the emission
measures into agreement with those from \aia.

The fits to the RHESSI spectra shown in Figure~\ref{fig:spectra} indicate somewhat higher
temperatures than we have derived from the \xrt\ and \aia\ filter ratios. These temperatures are
not directly comparable since the \rhessi\ spectra in this analysis are not spatially resolved and
the filter ratios are taken from a very small area along a flare loop. We have analyzed the
\xrt\ data for the entire field of view of the flare. We obtain a temperature of $\log T \lesssim
7.0$, again smaller than the 13\,MK observed with \rhessi, and a volume emission measure of $\log
\mathrm{EM} \lesssim 49.3$, which is about three orders of magnitude larger than what is observed
with \rhessi. These differences in the temperatures and the magnitudes of the emission measures
suggest a distribution of temperatures in the flare loops rather than an isothermal
plasma. Further, it suggests that the temperature is likely peaking at about 10\,MK and falling
rapidly at higher temperatures.

To estimate the differential emission measure (DEM) in the flare loop of interest we need to invert
the equation
\begin{equation}
  I_\lambda = \frac{1}{4\pi} \int \epsilon_\lambda(T)\xi(T)\,dT,
\end{equation}
where $\epsilon_\lambda(T)$ is the temperature response of the line or bandpass that we have
observed to have intensity $I_\lambda$ and $\xi(T) = n_e^2\,ds/dT$ is essentially the distribution
of temperatures in the loop. Note that we calculate the differential emission measure distribution
but, to facilitate comparisons with the emission measures computed from isothermal models, we plot
$\xi(T)\,dT$.

Inverting this equation requires that we utilize observations over a wide range of
temperatures. Ideally we would include \eis\ intensities in these calculations. While \xrt\ is
sensitive to very high temperature emission, \ion{Fe}{23} 263.760, and \ion{Fe}{24}
192.040\,\AA\ provide more localized constraints on the DEM inversion. Significant intensity,
however, was observed in these lines in only a few rasters. At most times we observe only background
noise. The brightest emission was observed in an exposure taken at 14:22:48\,UT and in
Figure~\ref{fig:dem} we show the DEM computed using the intensities of the several \eis\ emission
lines observed at that time. This DEM, which was calculated used the Monte Carlo Markov chain
(MCMC) algorithm of \citet{kashyap1998}, is broadly consistent with the isothermal models discussed
previously. It peaks at about 10\,MK at falls off at both lower and higher temperatures.

To compute the DEM at all other times when only \aia\ and \xrt\ are available we include ``pseudo
intensities'' of $0\pm50$\,erg~cm$^{-2}$~s$^{-1}$~sr$^{-1}$ for the \eis\ \ion{Fe}{23} 263.760, and
\ion{Fe}{24} 192.040\,\AA\ lines. This biases the DEM to lower values above 10\,MK and provides
some consistency with the weak signal observed in these lines at most times. Figure~\ref{fig:dem}
shows the DEM computed in this way at three times. The DEM calculated at 14:22:48\,UT is broadly
consistent with the \eis-only calculation at this time.

Two properties of the DEM are of particular interest: the temperature of the peak and the slope
from the peak to lower temperatures. The peak in the DEM will constrain the energy flux. Large
energy fluxes will produce high temperatures and it is clear from the observations that the amount
of plasma at 10--20\,MK is relatively small for this event, as might be expected if the energy
fluxes occur on a power-law distribution. The slope of the DEM away from the peak is also an
important constraint on multi-threaded modeling of the flare. Modeling the event as many
small-scale strands that are evolving independently will tend to broaden the DEM and these
observations limit that. To make such comparisons more quantitative we have calculated a power-law
slope for the EM using a function of the form $T^\alpha$ from the peak to $\log T = 6.2$. Since the
MCMC code provides a statistical ensemble of solutions, we fit each one individually and record the
median and standard deviation in the results. During the initial part of the flare the slopes are
very steep with $\alpha$ typically in the range of 5--6. During the later part of the event cooler
plasma is observed, the temperature broadens out, and $\alpha$ is typically in the range of
1--3. Because of the uncertainty in the DEM calculation at temperatures above the peak, we do not
consider the high-temperature slope.
\section{Summary and Discussion}

We have presented the analysis of a microflare observed with \iris, \rhessi, \aia, \xrt, and
\eis. In combination, these observations give us a comprehensive view of both the energy deposition
at the flare footpoints, including the properties of the non-thermal electrons and the response of
the transition region to the heating that they produce, and the high-temperature loops that form
there. At the high spatial and temporal resolution of \iris\ we observe that footpoint dynamics are
characterized by numerous, small-scale impulsive bursts typically lasting 60\,s or less. The
resulting distribution of footpoint intensities in the \ion{Si}{4} slit-jaw images follows a
power-law with an index of about -1.7. \iris\ observations in \ion{Si}{4}, \ion{C}{2}, \ion{Mg}{2},
and \ion{O}{1} allow us to follow the propagation of the energy deposited in the footpoints to
lower heights in the solar atmosphere. We observe a progressively weaker response to heating with
depth.

The observations from \xrt, \aia, and \eis\ show the formation of high-temperature loops rooted in
these footpoints. These loops have a relatively narrow range of temperatures peaked at about
10\,MK, consistent with DEMs computed for large flares \citep[e.g.,][]{warren2013}. The comparison
of \xrt\ and \rhessi\ temperatures and emission measures from the full field of view of the event
suggests that the temperature distribution falls very rapidly from 10\,MK to 13\,MK.

This analysis highlights the difficulty of combining observations with vastly different spatial
resolution. For example, the images derived from \rhessi\ clearly show hard X-ray emission in the
vicinity of the footpoint brightenings observed with \iris. It is not clear, however, if this hard
X-ray emission is limited to specific footpoints or covers most of the flare ribbon. The strong
correlation between the integrated hard X-ray emission and \iris\ footpoint intensities shown in
Figure~\ref{fig:rhessi} suggests pervasive hard X-ray emission along the flare ribbon, but more
observations are needed to confirm this point.

The comprehensive observations provided by \iris, \rhessi, \eis, \xrt, and \aia\ create an
extensive set of requirements for numerical simulations to reproduce. A successful model of this
event must account for
\begin{enumerate}
\item an increase in \ion{Si}{4} intensity of about $10^3$ over the background level that persist
  for 600\,s or more, with smaller increases for \ion{C}{2}, \ion{Mg}{2}, and \ion{O}{1},
\item red-shifts in \ion{Si}{4} that persist for 600\,s or more, with weaker red-shifts in
  \ion{C}{2} and \ion{Mg}{2}, and no Doppler shifts for \ion{O}{1},
\item multi-component line profiles for \ion{Si}{4}, \ion{C}{2}, and \ion{Mg}{2} with little
  emission at the rest wavelength for \ion{Si}{4} but a significant stationary component for
  \ion{C}{2} and \ion{Mg}{2},
\item DEMs that are sharply peaked near 10\,MK or less,
\item loop pressures of about $\log P = 16.9$ (in units of cm$^{-3}$\,K).
\end{enumerate}
We stress that the spectroscopic constraints must be met for an area corresponding to a single
\iris\ pixel, not the entire flare. 

We are currently working on a companion paper \citep{reep2016} that focuses on the one-dimensional
hydrodynamic modeling of this event with the HYDrodynamics and RADiation code (HYDRAD;
\citealt{bradshaw2003,bradshaw2013}). Perhaps the most significant challenge presented by these
data is the persistent redshifts observed in \ion{Si}{4} and \ion{C}{2}. Persistent red-shifts have
been observed with \iris\ for a number of events and are not an idiosyncrasy of the data that we
have analyzed \citep[see][]{brosius2015,brannon2015,sadykov2015,polito2016}.

It has been known for some time that ``chromospheric condensations'' accompany evaporative upflows
during impulsive flare heating \citep[e.g.,][]{ichimoto1984, zarro1988, canfield1990}. These
downflows, however, are predicted to dissipate on time-scales of 20--60\,s \citep{fisher1989},
making it difficult to simulate the flows that we observe in a single \iris\ pixel with a single
loop model. Numerous, small-scale threads appear to be necessary.


\acknowledgments This work was supported by NASA's Hinode project. The research leading to these
results has also received funding from the European Community’s Seventh Framework Programme
(FP7/2007-2013) under grant agreement no. 606862 (F-CHROMA) (PJAS). IRIS is a NASA Small Explorer
developed and operated by LMSAL with mission operations executed at NASA Ames Research center and
major contributions to downlink communications funded by the Norwegian Space Center through an ESA
PRODEX contract.  CHIANTI is a collaborative project involving George Mason University, the
University of Michigan (USA) and the University of Cambridge (UK).


\end{document}